\documentclass[journal,twoside]{IEEEtran}
\usepackage[english]{babel}
\usepackage{graphicx}
\usepackage{subfigure}
\usepackage{multicol}
\usepackage{setspace}
\usepackage{amsmath}
\usepackage{epstopdf}
\usepackage{fancyhdr}
\usepackage{indentfirst}
\usepackage{enumerate}
\usepackage{caption}
\usepackage{leftidx}
\usepackage{amssymb}
\usepackage{float}
\usepackage{breqn}
\usepackage{bbm}
\usepackage{booktabs}

\usepackage{mathrsfs}
\usepackage{cite}
\usepackage{multirow}
\usepackage{totcount}
\usepackage{algorithm}
\usepackage{algorithmic}
\usepackage{color}
\usepackage{makecell}
\usepackage{textcomp}
\usepackage{bm}
\usepackage{array}
\usepackage{colortbl}
\allowdisplaybreaks[4]
\begin{document}
	\title{LEO Satellite-Enabled Random Access with Large Differential Delay and Doppler Shift}
	\author{Boxiao Shen, ~\IEEEmembership{Student Member, ~IEEE}, Yongpeng Wu, ~\IEEEmembership{Senior Member, ~IEEE}, Wenjun Zhang, ~\IEEEmembership{Fellow, ~IEEE}, Symeon Chatzinotas, ~\IEEEmembership{Fellow, ~IEEE}, and Björn Ottersten, ~\IEEEmembership{Fellow, ~IEEE}
		\thanks{\emph{(Corresponding author: Yongpeng Wu)}}
		\thanks{This article was presented in part at the IEEE
			Globecom 2023 \cite{o1}.}
		\thanks{B. Shen, Y. Wu, and W. Zhang are with the Department of Electronic Engineering, Shanghai Jiao Tong
			University, Shanghai 200240, China (e-mails:
			\{boxiao.shen, yongpeng.wu, zhangwenjun\}@sjtu.edu.cn).
		}
		\thanks{
			S. Chatzinotas and B. Ottersten are with the Interdisciplinary Center for Security, Reliability
			and Trust (SnT), University of Luxembourg, 1855 Luxembourg City, Luxembourg (e-mails: 
			\{Symeon.Chatzinotas, bjorn.ottersten\}@uni.lu).
		}
	}
	
	\maketitle
	\thispagestyle{empty}
	\IEEEpeerreviewmaketitle
	\begin{abstract}
		This paper investigates joint device identification, channel estimation, and symbol detection for LEO satellite-enabled grant-free random access systems, specifically targeting scenarios where remote Internet-of-Things (IoT) devices operate without global navigation satellite system (GNSS) assistance. Considering the constrained power consumption of these devices, the large differential delay and Doppler shift are handled at the satellite receiver. We firstly propose a spreading-based multi-frame transmission scheme with orthogonal time-frequency space (OTFS) modulation to mitigate the doubly dispersive effect in time and frequency, and then analyze the input-output relationship of the system. Next, we propose a receiver structure based on three modules: a linear module for identifying active devices that leverages the generalized approximate message passing algorithm to eliminate inter-user and inter-carrier interference; a non-linear module that employs the message passing algorithm to jointly estimate the channel and detect the transmitted symbols; and a third module that aims to exploit the three dimensional block channel sparsity in the delay-Doppler-angle domain. Soft information is exchanged among the three modules by careful message scheduling. Furthermore, the expectation-maximization algorithm is integrated to adjust phase rotation caused by the fractional Doppler and to learn the hyperparameters in the priors. Finally, the convolutional neural network is incorporated to enhance the symbol detection. Simulation results demonstrate that the proposed transmission scheme boosts the system performance, and the designed algorithms outperform the conventional methods significantly in terms of the device identification, channel estimation, and symbol detection.
	\end{abstract}
	
	
	\section{Introduction}
	Internet-of-Things (IoT) is one of the most important scenarios for the next-generation communications\cite{c32}. A considerable part of IoT devices is distributed in remote areas to support applications, such as smart agriculture, climate monitoring, and intelligent transportation systems \cite{s41}. However, existing cellular networks, mainly located in densely populated regions, struggle to provide the ubiquitous connectivity required by these diverse IoT applications. Recently, low earth orbit (LEO) satellites have gained substantial interest, positioned to complement and expand existing terrestrial networks to achieve global connectivity\cite{c30}. The commercial interest in LEO satellite-enabled access first surged in the 1990s. Unfortunately, most of these early attempts failed, and only a few, such as Iridium, Globalstar, and Orbcomm, have had varying degrees of success\cite{c29}. Nowadays, the advancements in aerospace, production, and communication technologies have renewed the LEO satellite market, where systems like Starlink and OneWeb have deployed extensively\cite{c31}. These developments pave the way for LEO satellite communications to offer seamless global coverage. In contrast to traditional human-centric communications, IoT primarily relies on machine-type communications (MTC), which are often delay-tolerant, low data-rate and in some cases periodic\cite{c33}. In this scenario, the random access protocol is crucial for enabling efficient connectivity.
	
	Grant-free random access (GFRA) preferred in MTC has been proposed to reduce signaling overhead and enhance access capability \cite{c34}. Over the past few years, the extensive research has been conducted for joint device identification and channel estimation (JDICE) in terrestrial GFRA systems \cite{s32,s322,s33,s36,s44,s43,c9}. Specifically, \cite{s32,s322,s33} treated JDICE as a sparse signal recovery problem within the framework of compressed sensing. Here, the device pilots serve as a sensing matrix, and the problem is tackled using the approximate message passing (AMP) algorithm. The authors in \cite{s36} employed block sparse Bayesian learning to develop a low-complexity message passing (MP) solution. By exploiting the sparse feature and low-rank structure of the device state matrix, \cite{s44} proposed a dimension reduction to decrease the computational complexity and developed a Riemannian trust-region algorithm for JDICE. The integration of orthogonal frequency division multiplexing (OFDM) into GFRA systems was investigated in \cite{s43} and \cite{c9}. The authors in \cite{s43} proposed the generalized multiple measurement vector AMP algorithm to exploit the channel sparsity in the angular domain, with the state evolution framework provided to predict performance. The authors in \cite{c9} further addressed the challenges of timing and frequency offsets by modeling their effects as phase shifts on the pilot matrix, leading to a multiple measurement vector recovery problem for JDICE. Then, to effectively exploit the structured sparsity resulting from this model, a structured generalized approximate message passing (GAMP) algorithm was developed. To further improve system performance, \cite{Joint1} and \cite{Joint2} adopted spreading-based transmission schemes and designed an MP-based algorithm to perform joint device identification, channel estimation, and symbol detection (JDICESD). It should be noted that the schemes in \cite{s32,s322,s33,s36,s44,s43,c9,Joint1,Joint2} are designed for block fading channels, which are assumed to remain constant during one block transmission. Meanwhile, \cite{c9} does not take the significant Doppler shift into account and assumes that the frequency offset tends to zero. However, the inherent high mobility of LEO satellites introduces significant Doppler shifts, leading to rapid time variability on the terrestrial-satellite links (TSLs), probably resulting in outdated channel state information \cite{ss32}. Moreover, the long distance between the ground devices and satellites induces a larger propagation delay as compared to terrestrial systems. These issues may also cause severe inter-symbol and inter-carrier interference, degrading the performance of existing algorithms. Consequently, current terrestrial GFRA frameworks, without adjustments or new designs, cannot be directly applied to the highly dynamic LEO satellite IoT communications.
	
	The 3rd generation partnership project (3GPP) recommendations for non-terrestrial networks (NTN) address large delays and Doppler shifts depending on whether the terrestrial device is equipped with a built-in global navigation satellite system (GNSS) or not \cite{3gpp2,3gpp}. Devices with GNSS pre-compensate for Doppler shift and propagation delay before uplink transmission, utilizing the knowledge of satellite ephemeris and device's location. Conversely, for devices without GNSS, the satellite broadcasts a common delay during the initial cell-search procedure. Following this, the differential delay and Doppler shift are either pre-compensated at the device side or managed by the satellite during uplink transmission. This particular scenario is a critical focus in ongoing 3GPP NTN standardization efforts, with several work items actively seeking solutions \cite{o2,o3}. Notably, most existing traditional LEO satellite-enabled GFRA schemes \cite{c7,c6,c10,c5} are tailored primarily for the first scenario, where the residual delay and Doppler shift remain within a small range. Specifically, \cite{c7} and \cite{c6} assumed a perfect compensation: In \cite{c7}, the authors transformed the received signal to a tensor decomposition form and proposed a Bayesian learning algorithm for JDICE. Meanwhile, \cite{c6} investigated the joint device identification and data detection, where the active device maps the data to a codeword from a predetermined and unique codebook. Then, the active devices and the transmitted codewords were detected by maximizing the likelihood function of the received signal. Additionally, the authors in \cite{c10} proposed a OFDM-symbol repetition technique combined with a grid-based parametric model to address the residual delay and Doppler shift. Based on this model, a modified variance state propagation algorithm was designed for JDICE. Furthermore, \cite{c5} focused on the line-of-sight path, assuming asynchronous signals at the frame level but synchronous at the slot level, and developed an AMP-based algorithm for joint active device, delay, and Doppler detection. 
	
	However, as indicated in \cite{o2} and \cite{o3}, the GNSS-assisted schemes pose additional problems. Firstly, the procedure for updating satellite ephemeris at the device level remains ambiguous. Secondly, frequent estimation and compensation for Doppler and delay could dramatically reduce battery lifetime of the terrestrial devices. Remote IoT devices, specifically designed to be smaller, lighter, more powerful, and lower in cost, face particular challenges. For instance, devices in low-power wide-area networks (LPWANs) are expected to operate autonomously for about ten years with just two AA batteries \cite{o4}. Consequently, GNSS-based solutions may be impractical for these remote IoT applications. In scenarios without GNSS, a significant issue is the persistence of large differential delay and Doppler shift, with the differential delay potentially exceeding one symbol duration and the Doppler shift surpassing one subcarrier spacing. To tackle this issue, the authors in \cite{o2} proposed to pre-compensate the Doppler shift at the terrestrial device level using the initial cell-search procedure in the grant-based random access. For the differential delay problem, the inherent phase ambiguity problem is solved using a discrimination criterion based on the estimation of Doppler rate. Building on this, \cite{o3} introduced a refinement estimation stage utilizing change point detection. However, both schemes proposed in \cite{o2} and \cite{o3} still require terrestrial devices to pre-compensate for a major portion of Doppler shift, adversely affecting their energy efficiency. Orthogonal time frequency space (OTFS) modulation has emerged as a promising solution to ensure reliable communications in LEO satellite-enabled GRFA systems \cite{c35,c11,c4}, which converts the time-frequency channels into quasi-static delay-Doppler channels\cite{o5,c8}. In \cite{c35}, OTFS is combined with tandem spreading multiple access to accommodate the differential Doppler shift characteristics in LEO satellite-enabled GFRA. To exploit the spatial diversity, the integration of massive multiple-input multiple-output (MIMO) with OTFS-based GFRA has been explored in \cite{c11} and \cite{c4}. Specifically, \cite{c11} introduced a two-dimensional (2D) pattern coupled hierarchical prior within sparse Bayesian learning for JDICE, exploiting channel sparsity in the delay-Doppler-angle domain. \cite{c4} investigated OTFS aided by training sequences and proposed a two-stage JDICE scheme alongside a streamlined multi-user symbol detection method. It is worth noting that techniques in \cite{c35,c11,c4} require terrestrial devices to pre-compensate for a substantial part of the propagation delay, introducing additional complexity for remote IoT devices.
	
	Overall, previous studies have primarily focused on GNSS-based solutions, which are unsuitable for power-limited remote IoT applications due to their high energy demands. Even non-GNSS solutions, such as those in \cite{o2} and \cite{o3}, and recent OTFS-based GFRA systems, require terrestrial devices to pre-compensate for significant Doppler shift or delay, increasing their complexity and energy consumption. In this paper, we investigate LEO satellite-enabled GFRA systems where devices operate without GNSS. To further reduce energy consumption, we eliminate the need for terrestrial devices to pre-compensate for the large differential delay and Doppler shift. Instead, we propose a spreading-based multi-frame OTFS transmission scheme, combined with a MP-based JDICESD algorithm to mitigate these negative effects at the satellite receiver. Our main contributions are summarized as follows:
	\begin{itemize}
		\item We propose a spreading-based multi-frame OTFS transmission scheme to mitigate the doubly dispersive effect and increase the undersampling ratio by exploiting the quasi-static properties of the channel in the delay-Doppler domain. Subsequently, we conduct a detailed analysis of the symbol-level input-output relationship within the system.
		\item Based on the proposed transmission scheme, a MP-based algorithm is designed for JDICESD. We divide the receiver structure into three modules: Delay-wise device activity identification (DDAI) module handles the received signal along the delay dimension in parallel and employs the GAMP algorithm to eliminate the inter-user and inter-carrier interference; Joint channel estimation and symbol detection (JCESD) module addresses the nonlinear coupling of the activity state, channel coefficient, and transmit symbols of each device, where an MP algorithm is derived in a symbol-by-symbol fashion for JCESD; and the 3D sparsity exploitation (TSE) module aided by Markov random field aims to exploit the 3D block sparsity of channel in the delay-Doppler-angle domain. The soft information is exchanged among the three modules by carefully message scheduling. Furthermore, the expectation-maximization (EM) algorithm is embedded to accommodate the phase rotation caused by the fractional Doppler and to learn the hyperparameters in priors.
		\item The convolutional neural network (CNN) detector is specifically designed to exploit the statistical information provided by the proposed MP-based algorithm, thereby enhancing symbol detection. Particularly, the statistical information for each transmitted symbols can be extracted individually from the outputs of the MP-based algorithm. This capability enables the CNN detector to efficiently estimate all the transmitted symbols in parallel and adapt to the dynamic number of active devices.
		\item Comprehensive experiments are conducted to evaluate the performance of the proposed schemes. The results demonstrate that the proposed transmission scheme boosts system performance significantly. Notably, in the absence of fractional Doppler, the greater the number of transmitted frames, the better the performance. On the other hand, when the fractional Doppler is present, a moderate number of frames yields the best performance. Moreover, the simulation results indicate that the proposed MP-based algorithms outperform existing methods significantly in terms of device identification, channel estimation, and symbol detection; The CNN-enhanced detector further improves the symbol detection performance of the MP-based algorithm.
	\end{itemize}

	The rest of this paper is organized as follows. Section \ref{System Model} introduces the transmission scheme and formulates the problem. Section \ref{JDICESD Algorithm} proposes the MP-based algorithm for JDICESD. Section \ref{CNN-Enhanced Detector} design the CNN-enhanced detector. Section \ref{Numerical Results} evaluates the performance of the proposed schemes, followed by the conclusions in Section \ref{Conclusion}.
	
	\emph{Notations}: The superscripts $(\cdot)^{*}$ and $(\cdot)^{\mathrm{H}}$ denote the conjugate and conjugated-transpose operations, respectively. The boldface letters denote matrices or vectors. $\bar{\jmath}=\sqrt{-1}$ denotes the imaginary unit. $\operatorname{diag}(\mathbf{x})$ denotes the diagonal matrix with the elements of $\mathbf{x}$ on the main diagonal, and $\mathrm{vec}_{\text d}(\mathbf X)$ returns the main diagonal elements of a square matrix $\mathbf X$. $\|\mathbf{X}\|_{\text F}$ denotes the Frobenius norm of $\mathbf{X}$. $\lceil x\rceil$ denotes the smallest integer that is not less than $x$. $\odot$ is the Hadamard product operator and $\otimes$ denotes Kronecker product. $(\cdot)_{M}$ denotes mod $M$, and $\langle x\rangle_{N}$ denotes $\left(x+\left\lfloor\frac{N}{2}\right\rfloor\right)_{N}-\left\lfloor\frac{N}{2}\right\rfloor .$ The notation $\triangleq$ is used for definitions. $\delta(\cdot)$ denotes the Dirac delta function. $\mathbf{X}[a,:]$ denotes the $a$-th row of $\mathbf{X}$, wihle $\mathbf X[:,b]$ denotes the $b$-th column of $\mathbf{X}$. $X[a,b]$ and $x_{a,b}$ denote the $(a, b)$-th element of $\mathbf{X}$. $x_a$ denotes the $a$-th element of $\mathbf{x}$.
	
	\section{System Model}
	\label{System Model}
	\begin{figure}[!htb]
		\centering
		\captionsetup{font={small}}
		\setlength{\belowcaptionskip}{-.1cm}
		\includegraphics[width=3.5in]{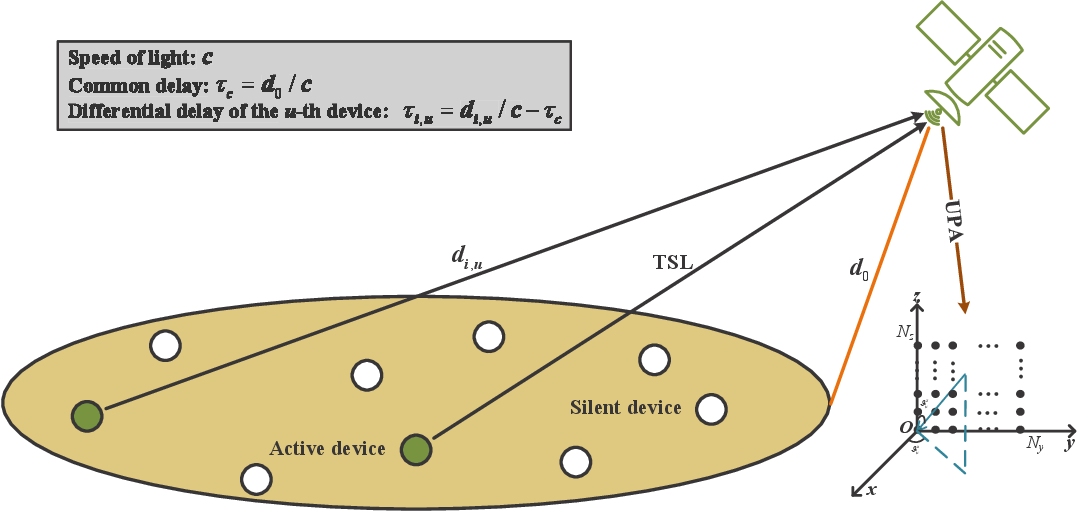}
		\caption{LEO satellite-enabled IoT network.}
		\label{system}
	\end{figure}
	We consider a LEO satellite-enabled grant-free random access system with the fixed beam as illustrated in Fig. \ref{system}, where OTFS is adopted to mitigate the doubly dispersive effect \cite{ss32} inherent in TSLs. The system comprises $U$ single-antenna terrestrial devices that communicate with a LEO satellite equipped with a uniform planar array (UPA) consisting of $N_a = N_z \times N_y$ antennas and a regenerative payload for on-board baseband signal processing. During each time interval, active devices share the same time-frequency resources to transmit their signals to the satellite. Additionally, to minimize energy consumption, we assume that these terrestrial devices operate without GNSS capabilities and do not pre-compensate for delay and Doppler shifts. Following 3GPP recommendations \cite{3gpp2,3gpp}, the satellite initially broadcasts a common delay for all devices, and subsequently handles the differential delay and Doppler shift observed in the uplink transmission.
	\subsection{Terrestrial-satellite Link Model}
	The TSLs will experience fast variations due to the high mobility of the LEO satellite. Hence, this time variation and
	multipath propagation require a doubly dispersive channel model for accurate TSLs characterization. Then, the delay-Doppler-space domain channel from the $u$-th device to the $(n_z+n_yN_z)$-th antenna at the satellite can be represented as \cite{ss32}
	\begin{align}
		\label{TSL}
		h_{u}^{n_z,n_y}(\tau, \nu)=\sum_{i=1}^{P} &h_{i,u} \delta\left(\tau-\tau_{i,u}\right) \delta\left(\nu-\nu_{i,u}\right) \nonumber \\
		&\times e^{\bar{\jmath} \pi n_z \Theta_{u}^z} 
		e^{\bar{\jmath} \pi n_y \Theta_{u}^y}.
	\end{align} 
	where $u = 0,\dots,U-1$, $n_z = 0,\dots,N_z-1$, $n_y = 0,\dots,N_y-1$, and $P$ is the number of physical paths. As shown in Fig. \ref{system}, $\tau_{i,u}=\tilde{\tau}_{i,u}-\tau^{\text c}$ represents the differential delay for the $i$-th path from the $u$-th device to the satellite, where $\tilde{\tau}_{i,u}$ and $\tau^{\text c}$ are denoted as propagation delay and common delay, respectively. The common delay corresponds to the propagation time between the satellite and the closest point on earth within the beam coverage \cite{o2}. Additionally, the notations $h_{i,u}$ and $\nu_{i,u}$ are the path gain and Doppler shift of the $i$-th path from $u$-th device to the satellite, respectively. The directional cosines $\Theta_{u}^z=\cos \vartheta_{u}^{\text z}$ and $\Theta_{u}^y = \sin \vartheta_{u}^{\text z} \sin \vartheta_{u}^{\text a}$, along the z-axis and y-axis, depend on the zenith angle $\vartheta_{u}^{\text z}$ and azimuth angle $\vartheta_{u}^{\text a}$, respectively. 
	
	Note that previous schemes based on the OTFS \cite{c35,c11,c4,o5,c8,s22,s20} generally require the subcarrier spacing $\Delta f > 2\nu_{\text{max}}$ and the symbol duration $T > \tau_{\text{max}}$ with $\Delta f =1/T$, where $\tau_{\text{max}}=\max\{\tau_{i,u}\}$ and $\nu_{\text{max}}=\max\{\left|\nu_{i,u}\right|\}$. In other words, these schemes are only suitable for the underspread channel that $\tau_{\text{max}}\nu_{\text{max}}<1$, which may not be satisfied in our considered scenario. For example, as computed in \cite{o2}, the differential delay and Doppler shift can be up to 698.62 \textmu s and $\pm 41$ kHz for the 3GPP Set-2, resulting in $\tau_{\text{max}}\nu_{\text{max}}\approx28.64$. As established in Proposition 1, the effective channel in this case is the 3D tensor at each antenna, which requires more observations to facilitate the channel estimation. This limitation motivates the development of our proposed transmission scheme. 
	\subsection{Transmission Scheme}
	\label{transmission scheme}
	The spreading-based multi-frame OTFS transmission scheme is proposed to handle the large differential delay and Doppler shift. In this scheme, each device is assigned an unique spreading code and its transmitted data symbols are spread into $Q$ consecutive frames during once transmission, where these frames are experienced OTFS modulation and demodulation independently, and are collected for further joint processing at the receiver. Specifically, in the $q$-th OTFS frame, $q=0,\dots,Q-1$, the $u$-th device is assigned a spreading code $C_{u,q}[k,l]$, $k=\lceil -N/2\rceil,\dots,\lceil N/2\rceil-1$, $l=0,\dots,M-1$, where $M$ and $N$ are the number of subcarriers and OFDM symbols within one OTFS frame, respectively. Then, the transmitted signal in the delay-Doppler domain is given by $X^{\text{DD}}_{u,q}[k,l]=C_{u,q}[k,l]t_u^{\text d}[l]$, where $t_u^{\text d}[l]$ represents the data symbol selected from a predefined alphabet $\mathcal{A} = \{a_1, \dots, a_{\left|\mathcal{A}\right|}\}$ with cardinality $\left|\mathcal{A}\right|$. Through the inverse symplectic finite Fourier transform (ISFFT) \cite{o5}, $X^{\text{DD}}_{u,q}[k,l]$ is mapped to $X^{\text{TF}}_{u,q}[n,m]$ in the time-frequency domain given by
	\begin{align}
		\label{xx}
		X^{\text{TF}}_{u,q}[n, m]=\frac{1}{\sqrt{M N}} \sum_{k=\lceil-N / 2\rceil}^{\lceil N / 2\rceil-1} \sum_{l=0}^{M-1} X^{\mathrm{DD}}_{u,q}[k, l] e^{-\bar{\jmath} 2 \pi\left(\frac{m l}{M}-\frac{n k}{N}\right)},
	\end{align}
	where $k=\lceil -N/2\rceil,\dots,\lceil N/2\rceil-1$, $l=0,\dots,M-1$. Next, the OFDM modulator converts $X^{\text{TF}}_{u,q}[n,m]$ to a continuous signal $s_{u}(t)$ using a transmit waveform $g_{\text{tx}}(t)$ as
	\begin{align}
		s_{u}(t)=\sum\limits_{q=0}^{Q-1}\sum\limits_{m=0}^{M-1} \sum\limits_{n=0}^{N-1} &X^{\mathrm{TF}}_{u,q}[n, m] e^{\bar{\jmath} 2 \pi m \Delta f \left(t-T_{\text{cp}}-(n+qN)T_{\mathrm{sym}}\right)} \nonumber \\
		&\times g_{\mathrm{tx}}\left(t-(n+qN) T_{\mathrm{sym}}\right),
	\end{align}
	where $T_{\text{cp}}$ is the time duration of the cyclic prefix (CP), and $T_{\mathrm{sym}}=T + T_{\text{cp}}$ is the time duration of a OFDM symbol with CP. Note that the CP is added in this step.
	
	Then, the signal $s_{u}(t)$ is transmitted over the TSL defined in (\ref{TSL}), and the received signal at the $(n_z+N_zn_y)$-th antenna is given by
	\begin{align}
		r_{u}^{n_z,n_y}(t)=\iint h_{u}^{n_z,n_y}(\tau, \nu) s_{u}(t-\tau) e^{j 2 \pi \nu(t-\tau)} d \tau d \nu,
	\end{align}
	where the noise is omitted to simplify notation. The received symbol $Y_{n_z,n_y,u}^{\mathrm{TFS,q}}[n, m]$ in the time-frequency-space domain are sampled from the cross-ambiguity function  $A_{g_{\mathrm{rx}},r}^{n_z,n_y,u}(t, f)$ as
	\begin{align}
		\label{YTF}
		Y_{n_z,n_y,u}^{\mathrm{TFS,q}}[n, m]=\left. A_{g_{\mathrm{rx}}, r}^{n_z,n_y,u}(t, f)\right|_{t=(n+qN) T_{\mathrm{sym}}, f=m \Delta f},
	\end{align}
	where $g_{\mathrm{rx}}(t)$ is denoted as the received waveform and
	$A_{g_{\mathrm{rx}}, r}^{n_z,n_y,u}(t, f) = \int g_{\mathrm{rx}}^{*}\left(t^{\prime}-t\right) r_{u}^{n_z,n_y}\left(t^{\prime}\right) e^{-\bar{\jmath} 2 \pi f\left(t^{\prime}-T_{\text{cp}}-t\right)} d t^{\prime}$.
	Finally, the symplectic finite Fourier transform (SFFT) \cite{o5} maps the symbols into the delay-Doppler-space domain as 
	\begin{align}
		\label{yys}
		Y_{n_z,n_y,u}^{\mathrm{DDS,q}}[k, l]=\frac{1}{\sqrt{MN}} \sum_{n=0}^{N-1} \sum_{m=0}^{M-1} Y_{n_z,n_y,u}^{\mathrm{TFS,q}}[n, m] e^{\bar{\jmath} 2 \pi\left(\frac{m l}{M}-\frac{n k}{N}\right)}.
	\end{align}
	We denote the delay and Doppler taps for the $i$-th path of the $u$-th device as follows
	\begin{align}
		\label{tap}
		\tau_{i,u}=\frac{l_{i,u} + b_{i,u}M}{M\Delta f}, \nu_{i,u} = \frac{k_{i,u} + \tilde{k}_{i,u} + d_{i,u}N}{NT_{\text{sym}}},
	\end{align}
	where $l_{i,u} = 0,\dots,M-1$ and $k_{i,u} = \lceil -N/2\rceil,\dots,\lceil N/2 \rceil -1$ represent the indexes of the delay and Doppler tap, respectively. $b_{i,u}$ and $d_{i,u}$ are the integers accounting for the parts beyond $T$ and $\Delta f$, respectively. $\tilde{k}_{i,u} \in (-\frac{1}{2},\frac{1}{2}]$ is the fractional Doppler shift. In addition, we denote the delay tap index set as $\mathcal{L}_u = [l_{1,u},\dots,l_{P,u}]$. The fractional delay is not considered here since the resolution of the sampling time is sufficient to approximate the path delays to the nearest sampling points in typical wide-band systems \cite{s20}.
	Based on (\ref{TSL})-(\ref{tap}), we can derive the input-output relationship of the system, as given in the following proposition.
	
	\textit{Proposition 1}: Given $T_{\text{cp}} \geq \tau_{\text{max}}$ and rectangular waveform $g_{\mathrm{tx}}(t)$ and $g_{\mathrm{rx}}(t)$, the input-output relationship in the delay-Doppler-space domain of this spreading-based multi-frame OTFS transmission scheme is represented as  
	\begin{align}
		\label{yDDS}
		Y_{n_z,n_y,u}^{\mathrm{DDS},q}[k, l] =&\sum_{l^{\prime}=0}^{M-1} \sum_{k^{\prime}=\lceil-N / 2\rceil}^{\lceil N / 2\rceil-1} H_{n_z,n_y,u}^{\mathrm{DDS}}\left[k^{\prime}, l^{\prime}, l\right]t_u^{\text d}\left[\left(l-l^{\prime}\right)_{M}\right] \nonumber\\
		&\times \phi_{u,q}[l^{\prime}] C_{u,q}\left[\left\langle k-k^{\prime}\right\rangle_{N},\left(l-l^{\prime}\right)_{M}\right],
	\end{align}   
	where 
	\begin{align}
		\label{HDDS}
		&H_{n_z,n_y,u}^{\mathrm{DDS}}\left[k^{\prime}, l^{\prime}, l\right]=\frac{1}{\sqrt N} \sum_{i=1}^{P}h_{i,u} e^{\bar{\jmath} 2 \pi \nu_{i,u}l T_s} e^{\bar{\jmath} \pi n_z \Theta_{u}^z} e^{\bar{\jmath} \pi n_y \Theta_{u}^y} \nonumber \\ 
		&\quad \quad \quad \quad \quad \times \Pi_N(k^{\prime}-NT_{\text{sym}}\nu_{i,u}) \delta\left(l^{\prime} T_{\mathrm{s}}-(\tau_{i,u})_T\right), \\
		&\phi_{u,q}[l^{\prime}]= \begin{cases}e^{\bar{\jmath} 2 \pi \tilde{k}_{i,u}q}, & l^{\prime} \in \mathcal{L}_u, l^{\prime}=l_{i,u} \\ 0, & l^{\prime} \notin \mathcal{L}_u\end{cases},\label{phiq}
	\end{align}
	$\Pi_N(x)\triangleq\frac{1}{\sqrt N} \sum_{i=0}^{N-1} e^{-\bar{\jmath} 2 \pi \frac{x}{N} i}$, $T_{\mathrm{s}}=\frac{1}{M\Delta f}$ is the sampling interval, and $l_{i,u}$ is selected from $\mathcal{L}_u$. 
	
	\emph{Proof}: Please see Appendix \ref{Appendix A}. $\hfill\blacksquare$
	
	Note that the relationship (\ref{yDDS}) aligns with those in \cite{c4,s20} when adopting a single-frame OTFS transmission with replacing $C_{u,q}[k,l]t_u^{\text d}[l]$ for $X^{\text{DD}}_{u,q}[k,l]$, provided $\Delta f > 2\nu_{\text{max}}$ and $T > \tau_{\text{max}}$. It is observed in (\ref{HDDS}) that the effective channel $H_{n_z,n_y,u}^{\mathrm{DDS}}\left[k^{\prime}, l^{\prime}, l\right]$ from the $u$-th device to the $(n_z+n_yN_z)$-th antenna is represented as a 3D tensor related to $l$, which is the received symbol position along the delay dimension. However, the condition $\tau_{\text{max}}\nu_{\text{max}} > 1$ precludes simplifications via methods similar to those in \cite{s22,s20}, which poses the significant challenges to the conventional pilot-based single OTFS frame transmission scheme \cite{c35,c11,c4,o5,c8,s22,s20}, particularly in achieving accurate channel estimation and symbol detection. Fortunately, $H_{n_z,n_y,u}^{\mathrm{DDS}}\left[k^{\prime}, l^{\prime}, l\right]$ remains approximately constant over $Q$ consecutive frames, benefited by the quasi-static property of (\ref{TSL}) in the delay-Doppler domain. This characteristic allows the proposed scheme to gather more observations, thereby facilitating the joint design of channel estimation and symbol detection. In practice, the quasi-static property may be disrupted when the number of transmitted frames becomes large. For instance, based on the simulation parameters in Sec. \ref{Numerical Results}, the Doppler resolution is approximately 195 Hz and the delay resolution is around 2.1 \textmu s. Given an initial elevation angle $50^{\text{o}}$ and $Q=8$ frames, the absolute variations of the Doppler shift and delay can be computed as $\Delta \nu\approx 11.7$ Hz and $\Delta \tau\approx 0.6$ \textmu s, which are less than the one-tenth and one-third of the Doppler and delay resolutions, respectively. This indicates that the channel in the delay-Doppler domain remains approximately quasi-static during the transmission of $Q=8$ frames. When $Q$ increases to 25, we find that $\Delta \nu \approx 36.5$ Hz and $\Delta \tau\approx 1.9$ \textmu s. In this case, $\Delta \nu$ is still less than the one-fifth of the Doppler resolution, while $\Delta \tau$ is below but approaches the delay resolution, indicating that the quasi-static nature of the channel may start to degrade. Therefore, in practical communications, smaller $Q$ values are preferred to maintain the quasi-static property of the channel. Furthermore, the phase rotation $\phi_{u,q}[l^{\prime}]$ caused by fractional Doppler accumulates across multiple frames and will be addressed carefully in our algorithm design.
	
	To exploit the sparsity in the angular domain, the two-dimensional discrete Fourier transform is applied along the space dimension to obtain the input-output relationship in the delay-Doppler-angle domain as
	\begin{align}
		\label{SUIO}
		&Y_{a_z,a_y,u}^{\mathrm{DDA},q}[k, l]\nonumber\\
		&=\frac{1}{\sqrt{N_zN_y}} \sum_{n_z=0}^{N_z-1}\sum_{n_y=0}^{N_y-1} Y_{n_z,n_y,u}^{\mathrm{DDS},q}[k, l] e^{-\bar{\jmath} 2 \pi (\frac{a_z n_z}{N_z}+\frac{a_y n_y}{N_y})} \nonumber\\
		&=\sum_{l^{\prime}=0}^{M-1} \sum_{k^{\prime}=\lceil-N / 2\rceil}^{\lceil N / 2\rceil-1} H_{a_z,a_y,u}^{\mathrm{DDA}}\left[k^{\prime}, l^{\prime}, l\right] t_u^{\text d}\left[\left(l-l^{\prime}\right)_{M}\right]  \nonumber\\
		&\quad\quad\quad\quad\quad \times \phi_{u,q}[l^{\prime}] C_{u,q}\left[\left\langle k-k^{\prime}\right\rangle_{N},\left(l-l^{\prime}\right)_{M}\right], 
	\end{align}
	where $H_{a_z,a_y,u}^{\mathrm{DDA}}[k^{\prime}, l^{\prime}, l]$ represents the effective channel in the delay-Doppler-angle domain given by
	\begin{align}
		\label{HDDA}
		&H_{a_z,a_y,u}^{\mathrm{DDA}}\left[k^{\prime}, l^{\prime}, l\right]=\nonumber \\
		&\frac{1}{\sqrt N} \sum_{i=1}^{P}h_{i,u} e^{\bar{\jmath} 2 \pi \nu_{i,u}l T_s} \delta\left(l^{\prime} T_{\mathrm{s}}-(\tau_{i,u})_T\right) \Pi_N(k^{\prime}-NT_{\text{sym}}\nu_{i,u}) \nonumber \\
		&\quad \quad \quad \times\Pi_{N_z}(a_z-N_z\Theta_{u}^z/2) \Pi_{N_y}(a_y-N_y\Theta_{u}^y/2) ,
	\end{align}
	where $a_y = 0,\dots,N_y-1$ and $a_z = 0,\dots,N_z-1$ are indexes along the angular domain. Therefore, we can find in (\ref{HDDA}) that $H_{a_z,a_y,u}^{\mathrm{DDA}}\left[k^{\prime}, l^{\prime}, l\right]$ has dominant elements only if $k^{\prime} \approx NT_{\text{sym}}\nu_{i,u}-b_{i,u}N$, $l^{\prime} \approx (\tau_{i,u})_T M \Delta f$, $a_y \approx N_y\Theta_{u}^y/2$, and $a_z \approx N_z\Theta_{u}^z/2$, which indicates the channel sparsity in the delay-Doppler-angle domain.
	\subsection{Problem Formulation}
	According to the superposition principle, the $q$-th received frame in the delay-Doppler-angle domain is given by
	\begin{align}
		\label{scalarIO}
		&Y_{a_z,a_y,q}^{\mathrm{DDA}}[k, l]
		=\sum_{u=0}^{U-1}\lambda_u Y_{a_z,a_y,u}^{\mathrm{DDA},q}[k, l] + Z_{a_z,a_y,q}^{\mathrm{DDA}}[k, l], 
	\end{align}
	where $\lambda_u\in \{0,1\}$ is the activity indicator of the $u$-th device and $Z_{a_z,a_y,q}^{\mathrm{DDA}}[k, l]\sim \mathcal{CN}(0,\sigma^2)$ is the noise. To facilitate the following analysis, we rewrite (\ref{scalarIO}) into the matrix form for each $q$ and $l$ as
	\begin{align}
		\label{qframe}
		\mathbf Y_q^l =  \mathbf C^l_q (\bm \Phi_q \otimes \mathbf I_N) (\mathbf T^l \otimes \mathbf I_N )(\bm \Lambda \otimes \mathbf{I}_{MN}) \tilde{\mathbf H}^l + \mathbf Z_q^l,
	\end{align}
	where $\mathbf Y^l_q$ and $\mathbf Z^l_q$ are with the $(k,a_z+N_za_y)$-th element $Y_{a_z,a_y,q}^{\mathrm{DDA}}[k, l]$ and $Z_{a_z,a_y,q}^{\mathrm{DDA}}[k, l]$, respectively. 
	$\mathbf C^l_q=[\mathbf C_{0,q}^l,\dots,\mathbf C_{U-1,q}^l]$, where $\mathbf C_{u,q}^l\in\mathbb{C}^{N\times MN}$ is the spreading code matrix of the $u$-th device at the delay dimension $l$; $\mathbf C_{u,q}^l$ is a block circulant matrix due to the 2D circular convolution in (\ref{SUIO}), 
	and its sub-matrix is the circulant matrix defined by $[C_{u,q}[0,(l-l^{\prime})_M],C_{u,q}[1,(l-l^{\prime})_M],\dots,C_{u,q}[-1,(l-l^{\prime})_M]]$. $\bm \Phi_q=\mathrm{diag}\{\bm \phi_q\}$, where $\bm \phi_q \in \mathbb{C}^{UM}$ with the $(uM+l^{\prime})$-th element $\phi_{u,q}[l^{\prime}]$, and $\mathbf{I}_{N} \in \mathbb{R}^{N\times N}$ is the identity matrix. $\mathbf T^l = \mathrm{diag}([\mathbf t_0^l, \dots, \mathbf t_{U-1}^l])$, where  $\mathbf t_u^l=[t_u^{\text d}[(l)_M],\dots,t_u^{\text d}[(l-M+1)_M]]$.
	$\bm \Lambda = \mathrm{diag}(\bm{\lambda})$, where $\bm{\lambda} = [\lambda_0,\cdots,\lambda_{U-1}]$. $\tilde{\mathbf H}^{l} \in \mathrm{C}^{UMN\times N_a}$ is with the $(uMN+l^{\prime}N+k^{\prime}, a_z+N_za_y)$-th element $H_{a_z,a_y,u}^{\mathrm{DDA}}\left[k^{\prime}, l^{\prime}, l\right]$.
	Next, we collect the $Q$ received frames into a single matrix as 
	\begin{align}
		\label{Qframe}
		\mathbf Y^l &=  \mathbf C^{\phi,l} (\mathbf T^l \otimes \mathbf I_N ) \mathbf H^l + \mathbf Z^l,
	\end{align} 
	where $\mathbf Y^l = \left[(\mathbf Y_0^l)^{\text T},\dots,(\mathbf Y_{Q-1}^l)^{\text T}\right]^{\text T}$, $\mathbf H^l = (\Lambda \otimes \mathbf{I}_{MN})  \tilde{\mathbf H}^l$,  $\mathbf C^{\phi,l} = \left[(\mathbf C_0^l (\bm \Phi_0 \otimes \mathbf I_N))^{\text T},\dots,(\mathbf C_{Q-1}^l(\bm \Phi_{Q-1} \otimes \mathbf I_N))^{\text T}\right]^{\text T}$. Compared to the single OTFS frame transmission scheme, $\mathbf C^{\phi,l}$ with undersampling ratio $\frac{Q}{UM}$ allows for more observations to enhance the algorithm performance. We aim to estimate $(\mathbf H, \mathbf t)$ through the maximum a posterior probability (MAP) principle given as
	\begin{align}
		\label{map}
		(\hat{\mathbf H}, \hat{\mathbf t})=\arg \max \limits_{(\mathbf H, \mathbf t)} p(\mathbf H, \mathbf t \mid \mathbf Y),
	\end{align}
	where $\mathbf Y = [\mathbf Y^0,\dots,\mathbf Y^{M-1}]$, $\mathbf H = [\mathbf H^0,\dots,\mathbf H^{M-1}]$,  $\mathbf t = [\mathbf t_0,\dots,\mathbf t_{U-1}]$, and $\mathbf t_u = [t_u^{\text d}[0],\dots,t_u^{\text d}[M-1]]$.
	
	Problem (\ref{map}) is generally non-convex and difficult to solve.
	MP algorithms could provide possible solutions.
	However, the sensing matrix $\mathbf C^{\phi,l}$ is underdetermined due to the phase rotation caused by the fractional Doppler, and the variables to be estimated in (\ref{map}) are all coupled together to form a bilinear function. Hence, the exact MP based on the sum-product rule is difficult to implement and the existing low complexity AMP-type algorithms cannot be applied directly. To address this issue, we are developing an iterative algorithm with a carefully designed receiver structure and sophisticated message updates. Additionally, the phase ambiguity problem \cite{Joint1} inevitably arise in (\ref{map}) since both channel and data symbols are unknown in the receiver. Common methods for combating this problem include differential coding or asymmetric constellation \cite{phase1}. In this work, we assume that the asymmetric constellation is adopted by ground devices. The coding-related technique is worth investigating in the future work.
	
	\section{JDICESD Algorithm}
	\label{JDICESD Algorithm}
	In this section, we resort to Bayesian method for designing JDICESD algorithm, where the receiver structure is divided into three modules and the soft information is exchanged iteratively among them by careful scheduling. In addition, EM algorithm is embedded to adjust the phase rotation caused by fractional Doppler and to learn hyperparameters in priors.
	\begin{figure}[!htb]
		\centering
		\captionsetup{font={small}}
		\includegraphics[width=3in]{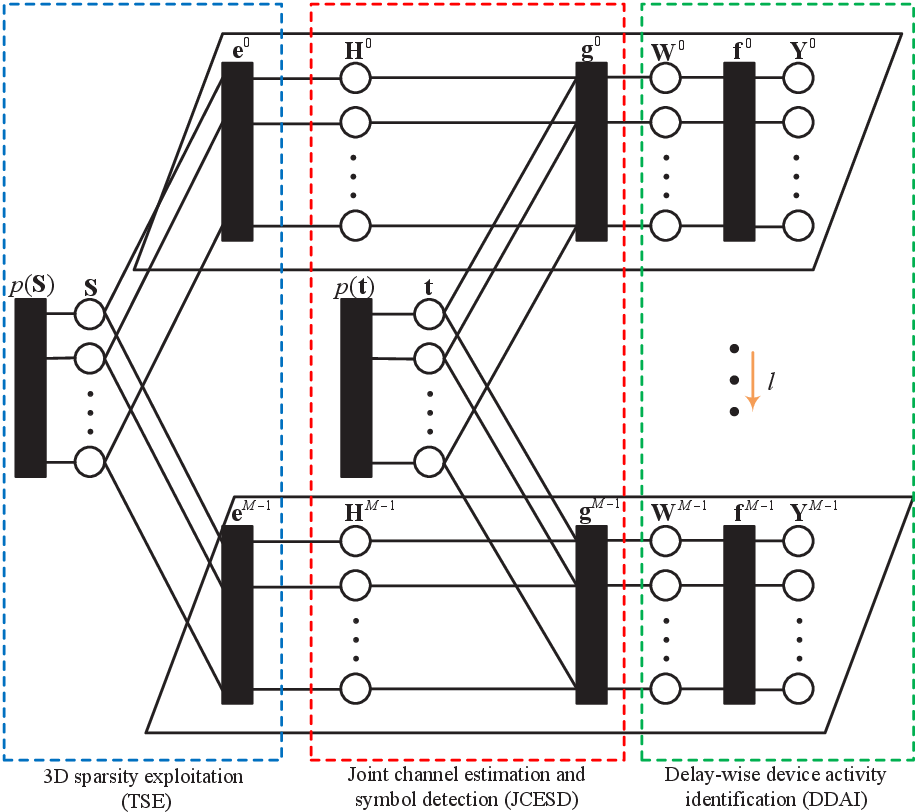}
		\caption{Factor graph representation.}
		\label{FactorGraph}
	\end{figure}
	\begin{figure*}[!htb]
		\centering
		\includegraphics[width=5in]{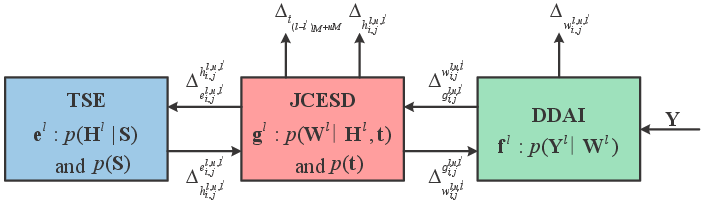}
		\caption{The block diagram of the inter-module communications.}
		\label{MC}
	\end{figure*}
	\subsection{Factor Graph Representation}
	Firstly, $\mathbf H^l$ is partitioned as sub-matrices $\mathbf H^{l,u}=\mathbf H^l[uMN:(u+1)MN-1,:] \in \mathbb{C}^{MN \times N_a}$ corresponding to the channel of the $u$-th device. Those sub-matrices are further split as $\mathbf H^{l,u,l^{\prime}} = \mathbf H^{l,u}[l^{\prime}N:(l^{\prime}+1)N-1,:] \in \mathbb{C}^{N \times Na}$, $l^{\prime}=0,\dots,M-1$, which is the channel of the $l^{\prime}$-th grid in the delay dimension of the $u$-th device. We denote the $(i,j)$-th element of $\mathbf H^{l,u,l^{\prime}}$ as $h^{l,u,l^{\prime}}_{i,j}$, $i=0,\cdots,N-1$, $j=0,\cdots,N_a-1$. Then, $K$-components Bernoulli-Gaussian-mixture (BGM) distribution is assigned for modeling $h^{l,u,l^{\prime}}_{i,j}$, which is a general and accurate distribution for the practical scenarios with the antenna array \cite{o6}, i.e,
	\begin{align}
		\label{hprior}
		&p(h_{i, j}^{l, u, l^{\prime}} \mid s_{i, j}^{l,u, l^{\prime}})= \delta(s_{i, j}^{l,u, l^{\prime}}+1) \delta(h_{i, j}^{l, u, l^{\prime}}) + \nonumber \\  &\quad \quad \quad \quad \quad \delta(s_{i, j}^{l,u, l^{\prime}}-1) \sum_{k=1}^K \omega_k^u \mathcal{CN}(h_{i, j}^{l, u, l^{\prime}}\mid \mu^u_k, \eta^u_k),  
	\end{align}  
	where $K$ and $(\omega_k^u, \mu^u_k, \eta^u_k)$ are the number of components and parameters of BGM, respectively, and $s_{i, j}^{l,u, l^{\prime}} \in \{+1,-1\}$ is the support of $h_{i, j}^{l, u, l^{\prime}}$. According to (\ref{HDDA}), the matrices $\{\mathbf H^{l,u,l^{\prime}}\}$ share the common support for each $l$, and hence we can let $s_{i, j}^{u, l^{\prime}}=s_{i, j}^{l,u, l^{\prime}}$. In addition, the channel sparsity in the delay-Doppler-angle domain turns into the 2D block sparsity in the matrix $\mathbf H^{l,u,l^{\prime}}$.  Then, we adopt the Markov random field (MRF) prior to describe the sparsity of $\mathbf H^{l,u,l^{\prime}}$, and the support can be characterized by the classic Ising model as
	\begin{align}
		\label{sprior}
		&p(\mathbf{S}^{u, l^{\prime}}) 
		\propto \exp \left(\sum_{i=0}^{N-1} \sum_{j=0}^{N_a-1}\left(\frac{1}{2} \sum_{s_{i^{\prime}, j^{\prime}}^{u, l^{\prime}} \in \mathcal D_{i, j}^{u, l^{\prime}}} \beta s_{i^{\prime}, j^{\prime}}^{u, l^{\prime}}-\alpha\right) s_{i, j}^{u, l^{\prime}}\right)
		\nonumber \\
		&\qquad=\left[\prod_{i,j}\prod_{s_{i^{\prime}, j^{\prime}} \in \mathcal D_{i, j}}  
		\psi(s_{i,j}^{u,l^{\prime}}, s_{i^{\prime}, j^{\prime}}^{u, l^{\prime}})\right]^{\frac{1}{2}}
		\prod_{i,j}\gamma(s_{i,j}^{u,l^{\prime}}),
	\end{align}
	where $\psi(s_{i,j}^{u,l^{\prime}}, s_{i^{\prime}, j^{\prime}}^{u, l^{\prime}}) = \exp(\beta s_{i, j}^{u, l^{\prime}}
	s_{i^{\prime}, j^{\prime}}^{u, l^{\prime}})$,
	$\gamma(s_{i,j}^{u,l^{\prime}}) = \exp(-\alpha s_{i, j}^{u, l^{\prime}})$,
	$\mathbf S^{u, l^{\prime}}$ is a support matrix with $(i,j)$-th element $s_{i, j}^{u, l^{\prime}}$, $\mathcal D_{i,j}^{u, l^{\prime}} = \{ s_{i-1,j}^{u,l^{\prime}}, s_{i+1,j}^{u,l^{\prime}}, s_{i,j-1}^{u,l^{\prime}}, s_{i,j+1}^{u,l^{\prime}} \}$ is the set containing the neighbors of $s_{i,j}^{u,l^{\prime}}$, and $\alpha$ and $\beta$ are the parameters of MRF prior; a larger $\beta$
	implies a larger size of each block of nonzeros, and a larger $\alpha$ encourages a sparser $\mathbf H^{l,u,l^{\prime}}$. To facilitate the algorithm design, we introduce the auxiliary variables 
	\begin{align}
		\label{auxiliary}
		\mathbf W^l = (\mathbf T^l \otimes \mathbf I_N ) \mathbf H^l, \quad \mathbf R^l=\mathbf C^{\phi,l}\mathbf W^l.
	\end{align}
	The similar splitting method can be adopted to get $\mathbf W^{l,u}$, $\mathbf W^{l,u,l^{\prime}}$, and $w^{l,u,l^{\prime}}_{i,j}$. We define $\mathbf W = [\mathbf W^0,\dots,\mathbf W^{M-1}]$ and  $\mathbf S = [\mathbf S^{0,0},\dots,\mathbf S^{U-1,M-1}]$. Then, the joint posterior distribution is given by
	\begin{align}
		\label{jointpost}
		&p(\mathbf W, \mathbf H, \mathbf t, \mathbf S \mid \mathbf Y) \nonumber\\
		&\propto \prod_l p(\mathbf Y^l \mid \mathbf W^l) p(\mathbf W^l \mid \mathbf H^l, \mathbf t) p(\mathbf t) p(\mathbf H^l \mid \mathbf S) p(\mathbf S) \nonumber \\
		&\propto \prod_l {\{}\prod_{o,j} p(y_{o,j}^l|r_{o,j}^l) \prod_{u,l^{\prime},i,j} {[} p(w^{l,u,l^{\prime}}_{i,j}|h_{i, j}^{l, u, l^{\prime}},t_{(l-l^{\prime})_M+uM})) \nonumber \\
		& \quad \quad \quad  p(h_{i, j}^{l, u, l^{\prime}} \mid s_{i, j}^{u, l^{\prime}}){]\}} 
		\prod_{u,l} p(t_{l+uM})) 
		\prod_{u,l^{\prime}} p(\mathbf{S}^{u, l^{\prime}}),
	\end{align}
	where 
	$p(w^{l,u,l^{\prime}}_{i,j}|h_{i, j}^{l, u, l^{\prime}},t_{(l-l^{\prime})_M+uM})= \delta(w^{l,u,l^{\prime}}_{i,j}-h_{i, j}^{l, u, l^{\prime}}t_{(l-l^{\prime})_M+uM})$ according to the equality constraint in (\ref{auxiliary}) and $p(y_{o,j}^l|r_{o,j}^l)=\mathcal{CN}(y_{o,j}^l|r_{o,j}^l,\sigma^2)$ due to the Gaussian noise.
	In addition, we assign the uniform distribution for the transmitted data symbols, i.e., $p(t_{l+uM})=\frac{1}{\left|\mathcal{A}\right|}\sum_{m=1}^{\left|\mathcal{A}\right|}\delta(t_{l+uM}-a_m)$.
	
	The factor graph representation of (\ref{jointpost}) is shown in Fig. \ref{FactorGraph}, where the factor nodes $\mathbf e^l$ with element $e_{i,j}^{l,u,l^{\prime}}$, $\mathbf g^l$ with element $g_{i,j}^{l,u,l^{\prime}}$, and $\mathbf f^l$ with element $f^l_{o,j}$ correspond to the distributions $p(h_{i, j}^{l, u, l^{\prime}} \mid s_{i, j}^{l,u, l^{\prime}})$, $p(w^{l,u,l^{\prime}}_{i,j}|h_{i, j}^{l, u, l^{\prime}},t_{(l-l^{\prime})_M+uM})$, and $p(y_{o,j}^l|r_{o,j}^l)$, respectively. 
	As shown in Fig. \ref{FactorGraph}, the receiver structure is divided into three modules, with the communication flow illustrated in Fig. \ref{MC}. Firstly, the soft message $\Delta_{w_{i,j}^{l,u,l^{\prime}}}^{g_{i,j}^{l,u,l^{\prime}}}$ from the JCESD module is initialized and serves as the prior for $\mathbf W^l$ in the DDAI module. The DDAI module, accounting for the factor $\mathbf f^l$, processes the received signal $\mathbf Y$ along with this soft message. It utilizes the GAMP algorithm to eliminate the inter-user interference in the delay-Doppler domain. This process yields the soft message $\Delta_{g_{i,j}^{l,u,l^{\prime}}}^{w_{i,j}^{l,u,l^{\prime}}}$, which is returned to the JCESD module as the likelihood for $\mathbf W^l$.
		Next, the JCESD module handles the non-linear factor $\mathbf g^l$. It takes $\Delta_{g_{i,j}^{l,u,l^{\prime}}}^{w_{i,j}^{l,u,l^{\prime}}}$as input and produces the soft message $\Delta^{h_{i,j}^{l,u,l^{\prime}}}_{e_{i,j}^{l,u,l^{\prime}}}$ as the likelihood for $\mathbf H^l$.
		Leveraging the MRF modeling $p(\mathbf S)$, the TSE module combines $\Delta^{h_{i,j}^{l,u,l^{\prime}}}_{e_{i,j}^{l,u,l^{\prime}}}$ with the factor $\mathbf e^l$ to exploit the 3D sparsity in $\mathbf H$. The refined prior $\Delta^{e_{i,j}^{l,u,l^{\prime}}}_{h_{i,j}^{l,u,l^{\prime}}}$ for $\mathbf H^l$ is then fed back to the JCESD module, which updates and provides the soft message $\Delta_{w_{i,j}^{l,u,l^{\prime}}}^{g_{i,j}^{l,u,l^{\prime}}}$ for the DDAI module. Subsequently, the DDAI module derives the estimated posterior distribution $\Delta_{w_{i,j}^{l,u,l^{\prime}}}$ for $\mathbf W$ by combining $\Delta_{w_{i,j}^{l,u,l^{\prime}}}^{g_{i,j}^{l,u,l^{\prime}}}$ and $\Delta_{g_{i,j}^{l,u,l^{\prime}}}^{w_{i,j}^{l,u,l^{\prime}}}$. These soft messages are iteratively exchanged among the three modules until convergence is achieved. Ultimately, the JCESD module integrates the soft messages $\Delta_{g_{i,j}^{l,u,l^{\prime}}}^{w_{i,j}^{l,u,l^{\prime}}}$, $\Delta^{h_{i,j}^{l,u,l^{\prime}}}_{e_{i,j}^{l,u,l^{\prime}}}$, and $\Delta^{e_{i,j}^{l,u,l^{\prime}}}_{h_{i,j}^{l,u,l^{\prime}}}$ to obtain the estimated posterior distributions $\Delta_{h_{i,j}^{l,u,l^{\prime}}}$ for $\mathbf H$ and $\Delta_{t_{(l-l^{\prime})_M+uM}}$ for transmitted symbols $\mathbf t$. Simultaneously, the approximated MMSE estimations for $\mathbf H$ and $\mathbf W$ are obtained as the means of their respective estimated posterior distributions. Then, the active devices are identified using the energy detector, given by
		\begin{align}
			\label{detectA}
			\hat{\lambda}_u = \mathbb{I}\left\{\sum_{l=0}^{M-1} \Vert\mathbf W^{l,u}\Vert_{\text F}^2 > \mathcal{T}\right\}, 
		\end{align}
		where $\mathbb{I}\{\cdot\}$ is the indicator function and $\mathcal{T}$ is the empirical predefined threshold. Moreover, the transmitted symbols are detected separately using $\Delta_{t_{(l-l^{\prime})_M+uM}}$ based on the MAP rule. In the following subsection, we will derive the soft messages within the receiver in detail.
	
	\subsection{Message Scheduling}
	In this subsection, we assume that the phase rotation caused by fractional Doppler, the noise variance, and the parameters of BGM are known, which will be updated later. In addition, $\Delta^{x}_{x^{\prime}}$ and $\Delta_{x}^{x^{\prime}}$ are denoted as the message passed from node $x$ to $x^{\prime}$ and $x^{\prime}$ to $x$, respectively. Now, we describe how the messages iterate among the three modules to get the final estimation. Firstly, the DDAI module aims to estimate $\mathbf W^l$ in the linear model, and hence the output message can be approximated by the GAMP\cite{sss}, given as
	\begin{align}
		\label{gampout}
		\Delta_{g_{i,j}^{l,u,l^{\prime}}}^{w_{i,j}^{l,u,l^{\prime}}}
		=\mathcal{CN}(w_{i,j}^{l,u,l^{\prime}}\mid\hat{r}_{i,j}^{w^{l,u,l^{\prime}}}, \tau_{i,j}^{w^{l,u,l^{\prime}}}),
	\end{align}
	where the mean $\hat{r}_{i,j}^{w^{l,u,l^{\prime}}}$ and the variance $\tau_{i,j}^{w^{l,u,l^{\prime}}}$ are updated iteratively by GAMP (see line 9 and 10 of Algorithm \ref{MRF-MP-GAMP}). Next, we focus on the messages scheduling of the JCESD and the TSE module.
	Combined with the output of DDAI module and the message from variable node $\mathbf t$ to check node ${\mathbf g^l}$, the message from ${\mathbf g^l}$ to $\mathbf H^l$ will be a BGM distribution, given as
	\begin{align}
		\label{CCESDfirst}
		&\Delta^{g_{i,j}^{l,u,l^{\prime}}}_{h_{i,j}^{l,u,l^{\prime}}}
		\propto 
		\int_{\sim h_{i,j}^{l,u,l^{\prime}}} 
		\Delta_{g_{i,j}^{l,u,l^{\prime}}}^{t_{(l-l^{\prime})_M+uM}} \Delta_{g_{i,j}^{l,u,l^{\prime}}}^{w_{i,j}^{l,u,l^{\prime}}}
		\times \nonumber \\
		&\qquad\qquad\qquad\qquad \delta(w_{i,j}^{l,u,l^{\prime}} - h_{i,j}^{l,u,l^{\prime}}t_{(l-l^{\prime})_M+uM}) \nonumber \\
		&=\sum_{m=1}^{|\mathcal A|} \overleftarrow{p}^{l,u,l^{\prime}}_{m,i,j} \mathcal{CN}(h_{i,j}^{l,u,l^{\prime}}a_m\mid \hat{r}_{i,j}^{w^{l,u,l^{\prime}}}, \tau_{i,j}^{w^{l,u,l^{\prime}}}),
	\end{align}
	where  $\Delta_{g_{i,j}^{l,u,l^{\prime}}}^{t_{(l-l^{\prime})_M+uM}} = \sum_{m=1}^{|\mathcal{A}|}
	\overleftarrow{p}^{l,u,l^{\prime}}_{m,i,j} \delta(t_{(l-l^{\prime})_M+uM}-a_m)$ is initialized as the uniform distribution over $\mathcal A$ and updated later. Combining $\Delta^{h_{i,j}^{l,u,l^{\prime}}}_{e_{i,j}^{l,u,l^{\prime}}}=\Delta^{g_{i,j}^{l,u,l^{\prime}}}_{h_{i,j}^{l,u,l^{\prime}}}$ with the conditional PDF of $h_{i,j}^{l,u,l^{\prime}}$ in (\ref{hprior}), the message from check node $\mathbf e^l$ to variable node $\mathbf S^{u,l^{\prime}}$ is the Bernoulli distribution, given as
	\begin{align}
		\label{rho}
		&\Delta^{e_{i,j}^{l,u,l^{\prime}}}_{s_{i,j}^{u,l^{\prime}}}
		\propto \int_{\sim s_{i,j}^{u,l^{\prime}}}
		p\left(h_{i, j}^{l, u, l^{\prime}} \mid s_{i, j}^{u, l^{\prime}} \right) \Delta^{g_{i,j}^{l,u,l^{\prime}}}_{h_{i,j}^{l,u,l^{\prime}}} \nonumber \\
		&=\rho_{i,j}^{l,u,l^{\prime}} \delta\left(s_{i, j}^{u, l^{\prime}}-1\right) + (1-\rho_{i,j}^{l,u,l^{\prime}})\delta\left(s_{i, j}^{u, l^{\prime}}+1\right),
	\end{align}
	where $\rho_{i,j}^{l,u,l^{\prime}}=\frac{\rho_{i,j}^{\text{A}^{l,u,l^{\prime}}}}{\mathcal{CN}(0\mid \hat{r}_{i,j}^{w^{l,u,l^{\prime}}}, \tau_{i,j}^{w^{l,u,l^{\prime}}}) + \rho_{i,j}^{\text{A}^{l,u,l^{\prime}}}}$, and 
	the auxiliary variable $\rho_{i,j}^{\text{A}^{l,u,l^{\prime}}}$ is defined as
	\begin{align}
		\label{yitaA}
		&\rho_{i,j}^{\text{A}^{l,u,l^{\prime}}}=\sum_{k=1}^K \sum_{m=1}^{|\mathcal A|}\omega_k^u \overleftarrow{p}^{l,u,l^{\prime}}_{m,i,j} \times \nonumber \\
		&\qquad \quad\mathcal{CN}\left(0\mid \hat{r}_{i,j}^{w^{l,u,l^{\prime}}}-\mu^u_k a_m, \tau_{i,j}^{w^{l,u,l^{\prime}}}+\eta^u_k \left|a_m\right|^2 \right).
	\end{align}
	With the inputs $\Delta^{e_{i,j}^{l,u,l^{\prime}}}_{s_{i,j}^{u,l^{\prime}}}(s_{i,j}^{u,l^{\prime}})$, we are now ready to describe the
	messages involved in the TSE module. To clearly characterize
	the relative position, the left, right, top, and bottom neighbors of $s_{i,j}^{u,l^{\prime}}$ are reindexed by $\{s_{i,j_{\text L}}^{u,l^{\prime}}, s_{i,j_{\text R}}^{u,l^{\prime}}, s_{i,j_{\text T}}^{u,l^{\prime}}, s_{i,j_{\text B}}^{u,l^{\prime}} \}$ corresponding to $\mathcal{D}_{i,j}^{u,l^{\prime}}$. The left, right, top, and bottom input messages of $s_{i,j}^{u,l^{\prime}}$,  denoted as $\Omega^{\text{L}^{u,l^{\prime}}}_{i,j}$, $\Omega^{\text{R}^{u,l^{\prime}}}_{i,j}$,$\Omega^{\text{T}^{u,l^{\prime}}}_{i,j}$, and $\Omega^{\text{B}^{u,l^{\prime}}}_{i,j}$, are Bernoulli distributions. $\Omega^{\text{L}^{u,l^{\prime}}}_{i,j}$ is given by
	\begin{align}
		\label{kexi}
		&\Omega^{\text{L}^{u,l^{\prime}}}_{i,j}
		\propto \nonumber \\
		&\int_{\sim s_{i,j}^{u,l^{\prime}}} \prod_{l=0}^{M-1}
		\Delta^{e_{i,j}^{l,u,l^{\prime}}}_{s_{i,j}^{u,l^{\prime}}}(s_{i,j}^{u,l^{\prime}}) \prod_{p \in \{\text L,\text T,\text B\}} 
		\Omega^{p^{u,l^{\prime}}}_{i,j_{\text{L}}} \gamma(s_{i,j_{\text L}}^{u,l^{\prime}}) 
		\psi(s_{i,j}^{u,l^{\prime}}, s_{i,j_{\text L}}^{u,l^{\prime}}) 
		\nonumber \\
		&=\xi^{\text{L}^{u,l^{\prime}}}_{i,j} \delta\left(s_{i, j}^{u, l^{\prime}}-1\right) + (1-\xi^{\text{L}^{u,l^{\prime}}}_{i,j}) \delta\left(s_{i, j}^{u, l^{\prime}}+1\right),
	\end{align}
	where $\xi^{\text{L}^{u,l^{\prime}}}_{i,j}$ is given in (\ref{xiupdate}).
	\begin{figure*}[!t]
		\normalsize
		\setcounter{equation}{26}
		\begin{equation}
			\label{xiupdate}
			\xi^{\text{L}^{u,l^{\prime}}}_{i,j} = \frac{e^{-\alpha+\beta}
				\prod_{l=0}^{M-1} \eta_{i,j_{\text L}}^{l,u,l^{\prime}}
				\prod_{p \in \{\text L,\text T,\text B\}} \xi^{p^{u,l^{\prime}}}_{i,j_{\text L}}	
				+ 
				e^{\alpha-\beta}
				\prod_{l=0}^{M-1} (1-\eta_{i,j_{\text L}}^{l,u,l^{\prime}})
				\prod_{p \in \{\text L,\text T,\text B\}} (1-\xi^{p^{u,l^{\prime}}}_{i,j_{\text L}})}{(e^{\beta}+e^{-\beta})
				\left(e^{-\alpha}\prod_{l=0}^{M-1} \eta_{i,j_{\text L}}^{l,u,l^{\prime}}
				\prod_{p \in \{\text L,\text T,\text B\}} \xi^{p^{u,l^{\prime}}}_{i,j_{\text L}}+
				e^{\alpha}
				\prod_{l=0}^{M-1} (1-\eta_{i,j_{\text L}}^{l,u,l^{\prime}})
				\prod_{p \in \{\text L,\text T,\text B\}} (1-\xi^{p^{u,l^{\prime}}}_{i,j_{\text L}})\right)} 
		\end{equation}
		\vspace{-0.4cm}
	\end{figure*}
	The input messages of $s_{i,j}^{u,l^{\prime}}$ from right, top, and bottom have a similar form to $\Omega^{\text{L}^{u,l^{\prime}}}_{i,j}$. Then, the output message from $s_{i,j}^{u,l^{\prime}}$ is given by
	\begin{align}
		&\Delta^{s_{i,j}^{u,l^{\prime}}}_{e_{i,j}^{l,u,l^{\prime}}} 
		\propto \gamma(s_{i,j}^{u,l^{\prime}}) \prod_{\hat l \neq l} \Delta^{e_{i,j}^{l,u,l^{\prime}}}_{s_{i,j}^{u,l^{\prime}}}
		\prod_{p\in\{\text L, \text R, \text T, \text B\}} \Omega^{p^{u,l^{\prime}}}_{i,j} \nonumber \\
		&\quad = \zeta_{i,j}^{l,u,l^{\prime}} \delta\left(s_{i, j}^{u, l^{\prime}}-1\right) +
		(1 - \zeta_{i,j}^{l,u,l^{\prime}}) \delta\left(s_{i, j}^{u, l^{\prime}}+1\right),
	\end{align}
	where $\zeta_{i,j}^{l,u,l^{\prime}}$ is given in (\ref{zitaupdate}).
	\begin{figure*}[!t]
		\normalsize
		\setcounter{equation}{28}
		\begin{equation}
			\label{zitaupdate}
			\zeta_{i,j}^{l,u,l^{\prime}} = 
			\frac{e^{-\alpha}\prod_{\hat l \neq l} \eta_{i,j}^{\hat l,u,l^{\prime}} \prod_{p\in\{\text L, \text R, \text T, \text B\}} \xi^{p^{u,l^{\prime}}}_{i,j}}{e^{-\alpha}\prod_{\hat l \neq l} \eta_{i,j}^{\hat l,u,l^{\prime}} \prod_{p\in\{\text L, \text R, \text T, \text B\}} \xi^{p^{u,l^{\prime}}}_{i,j}
				+ e^{\alpha}\prod_{\hat l \neq l} (1 - \eta_{i,j}^{\hat l,u,l^{\prime}}) \prod_{p\in\{\text L, \text R, \text T, \text B\}} (1-\xi^{p^{u,l^{\prime}}}_{i,j} )} 
		\end{equation}
		\vspace{-0.4cm}
	\end{figure*}
	The refined messages of $\mathbf H^l$ after exploiting sparsity will be fed back to the JCESD module, given as
	\begin{align}
		\label{TSEend}
		&\Delta^{e_{i,j}^{l,u,l^{\prime}}}_{h_{i,j}^{l,u,l^{\prime}}} 
		\propto \int_{s_{i,j}^{u,l^{\prime}}} \Delta^{s_{i,j}^{u,l^{\prime}}}_{e_{i,j}^{l,u,l^{\prime}}}
		p(h_{i, j}^{l, u, l^{\prime}} \mid s_{i, j}^{u, l^{\prime}})
		\nonumber \\
		&= (1 - \zeta_{i,j}^{l,u,l^{\prime}}) 
		\delta(h_{i,j}^{l,u,l^{\prime}}) + 
		\zeta_{i,j}^{l,u,l^{\prime}} \sum_{k=1}^K \omega_k^u \mathcal{CN}\left(h_{i, j}^{l, u, l^{\prime}}\mid \mu^u_k, \eta^u_k\right), 
	\end{align}
	Since $\Delta^{e_{i,j}^{l,u,l^{\prime}}}_{h_{i,j}^{l,u,l^{\prime}}}= \Delta^{h_{i,j}^{l,u,l^{\prime}}}_{g_{i,j}^{l,u,l^{\prime}}}$, the messages from $\mathbf g$ to $\mathbf t$ is given by
	\begin{align}
		\label{pright}
		&\Delta^{g_{i,j}^{l,u,l^{\prime}}}_{t_{(l-l^{\prime})_M+uM}} \propto \nonumber \\
		&\int_{h_{i,j}^{l,u,l^{\prime}},w_{i,j}^{l,u,l^{\prime}}}
		\Delta^{h_{i,j}^{l,u,l^{\prime}}}_{g_{i,j}^{l,u,l^{\prime}}}  \Delta_{g_{i,j}^{l,u,l^{\prime}}}^{w_{i,j}^{l,u,l^{\prime}}}\delta(w_{i,j}^{l,u,l^{\prime}} - h_{i,j}^{l,u,l^{\prime}}t_{(l-l^{\prime})_M+uM}) \nonumber \\
		&= \sum_{m=1}^{|\mathcal{A}|} 
		\overrightarrow{p}^{l,u,l^{\prime}}_{m,i,j} \delta(t_{(l-l^{\prime})_M+uM}-a_m),	
	\end{align}
	where $\overrightarrow{p}^{l,u,l^{\prime}}_{m,i,j}$ is given as
	\begin{align}
		&\overrightarrow{p}^{l,u,l^{\prime}}_{m,i,j} 
		\propto
		(1-\zeta_{i,j}^{l,u,l^{\prime}}) \mathcal{CN}\left(0\mid \hat{r}_{i,j}^{w^{l,u,l^{\prime}}}, \tau_{i,j}^{w^{l,u,l^{\prime}}}\right)+ \nonumber \\
		&\zeta_{i,j}^{l,u,l^{\prime}} \sum_{k=1}^K \omega_k^u 
		\mathcal{CN}\left(0\mid \hat{r}_{i,j}^{w^{l,u,l^{\prime}}}-\mu^u_k a_m, \tau_{i,j}^{w^{l,u,l^{\prime}}}+\eta^u_k \left|a_m\right|^2\right).	\nonumber	
	\end{align}
	Combined with the input messages to $t_{(l-l^{\prime})_M+uM}$, the output message from $t_{(l-l^{\prime})_M+uM}$ is given by
	\begin{align}	\Delta^{t_{(l-l^{\prime})_M+uM}}_{g_{i,j}^{l,u,l^{\prime}}} &\propto
		(\prod_{\substack{0\leq b<(l-l^{\prime})_M \\ b\neq l}} \prod_{i^{\prime}=0}^{N-1} \prod_{j^{\prime}=0}^{N_a-1} \Delta^{g_{i^{\prime},j^{\prime}}^{b,u,b+M-(l-l^{\prime})_M}}_{t_{(l-l^{\prime})_M+uM}}) \nonumber \\
		&\times(\prod_{\substack{(l-l^{\prime})_M\leq b< M\\ b\neq l}} \prod_{i^{\prime}=0}^{N-1} \prod_{j^{\prime}=0}^{N_a-1} \Delta^{g_{i^{\prime},j^{\prime}}^{b,u,b-(l-l^{\prime})_M}}_{t_{(l-l^{\prime})_M+uM}}) \nonumber \\
		&\times(\prod_{i^{\prime} \neq i}\prod_{j^{\prime}=0}^{N_a-1} \Delta^{g_{i^{\prime},j^{\prime}}^{l,u,l^{\prime}}}_{t_{(l-l^{\prime})_M+uM}}) 
		(\prod_{j^{\prime}\neq j} \Delta^{g_{i,j^{\prime}}^{l,u,l^{\prime}}}_{t_{(l-l^{\prime})_M+uM}}) \nonumber \\
		&= \sum_{m=1}^{|\mathcal{A}|}
		\overleftarrow{p}^{l,u,l^{\prime}}_{m,i,j} \delta(t_{(l-l^{\prime})_M+uM}-a_m).
	\end{align}
	Then, $\overleftarrow{p}^{l,u,l^{\prime}}_{m,i,j}$ can be updated by
	\begin{align}
		\label{pleft}
		\overleftarrow{p}^{l,u,l^{\prime}}_{m,i,j}
		= &\frac{1}{F_{i,j}^{l,u,l^{\prime}}}(\prod_{\substack{0\leq b<(l-l^{\prime})_M \\ b\neq l}} \prod_{i^{\prime}=0}^{N-1} \prod_{j^{\prime}=0}^{N_a-1} \overrightarrow{p}^{b,u,b+M-(l-l^{\prime})_M}_{m,i^{\prime},j^{\prime}}) \nonumber \\
		&\times (\prod_{\substack{(l-l^{\prime})_M\leq b\leq M\\ b\neq l}} \prod_{i^{\prime}=0}^{N-1} \prod_{j^{\prime}=0}^{N_a-1} \overrightarrow{p}^{b,u,b-(l-l^{\prime})_M}_{m,i^{\prime},j^{\prime}}) \nonumber\\
		&\times( \prod_{i^{\prime} \neq i} \prod_{j^{\prime}=0}^{N_a-1} \overrightarrow{p}^{l,u,l^{\prime}}_{m,i^{\prime},j^{\prime}})
		(\prod_{j^{\prime}\neq j} \overrightarrow{p}^{l,u,l^{\prime}}_{m,i,j^{\prime}})
	\end{align}
	where $F_{i,j}^{l,u,l^{\prime}}$ is the normalization constant such that $\sum_{m=1}^{|\mathcal A|}\overleftarrow{p}^{l,u,l^{\prime}}_{m,i,j}=1$. Next, given $\Delta^{h_{i,j}^{l,u,l^{\prime}}}_{g_{i,j}^{l,u,l^{\prime}}}$ and $\Delta^{t_{(l-l^{\prime})_M+uM}}_{g_{i,j}^{l,u,l^{\prime}}}$, the message of feedback from JCESD module to DDAI is the BGM distribution, given by
	\begin{align}
		&\Delta_{w_{i,j}^{l,u,l^{\prime}}}^{g_{i,j}^{l,u,l^{\prime}}}
		\propto \nonumber \\
		&\int_{\sim w_{i,j}^{l,u,l^{\prime}}} \delta(w_{i,j}^{l,u,l^{\prime}} - h_{i,j}^{l,u,l^{\prime}}t_{(l-l^{\prime})_M+uM})
		\Delta^{h_{i,j}^{l,u,l^{\prime}}}_{g_{i,j}^{l,u,l^{\prime}}}
		\Delta^{t_{(l-l^{\prime})_M+uM}}_{g_{i,j}^{l,u,l^{\prime}}}
		\nonumber \\
		&=(1-\zeta_{i,j}^{l,u,l^{\prime}})\delta(w_{i,j}^{l,u,l^{\prime}}) + \zeta_{i,j}^{l,u,l^{\prime}} \times \nonumber \\
		&\sum_{k=1}^K \sum_{m=1}^{|\mathcal A|}\omega_k^u \overleftarrow{p}^{l,u,l^{\prime}}_{m,i,j}
		\mathcal{CN}\left(w_{i,j}^{l,u,l^{\prime}}\mid \mu^u_k a_m, \eta^u_k \left|a_m\right|^2\right), 
	\end{align}
	Now, the posterior distribution of $w_{i,j}^{l,u,l^{\prime}}$ is approximated as a BGM distribution by combing all the input messages to it, given by
	\begin{align}
		\label{postw}
		&\Delta_{w_{i,j}^{l,u,l^{\prime}}}
		\propto
		\Delta_{w_{i,j}^{l,u,l^{\prime}}}^{g_{i,j}^{l,u,l^{\prime}}} \Delta_{g_{i,j}^{l,u,l^{\prime}}}^{w_{i,j}^{l,u,l^{\prime}}} 
		=(1 - \chi_{i,j}^{l,u,l^{\prime}})\delta(w_{i,j}^{l,u,l^{\prime}})
		+ \chi_{i,j}^{l,u,l^{\prime}} \nonumber \\
		&\times \sum_{k=1}^{K}\sum_{m=1}^{\mathcal |A|}\bar{\omega}_{k,m,i,j}^{l,u,l^{\prime}}\mathcal{CN}(w_{i,j}^{l,u,l^{\prime}}\mid \bar{\theta}_{k,m,i,j}^{l,u,l^{\prime}}, \bar{\varphi}_{k,m,i,j}^{l,u,l^{\prime}}),
	\end{align}
	where 
	\begin{align}
		\label{poststart}
		&\chi_{i,j}^{l,u,l^{\prime}} = 
		\frac{\zeta_{i,j}^{l,u,l^{\prime}}\rho_{i,j}^{\text{A}^{l,u,l^{\prime}}}}{(1-\zeta_{i,j}^{l,u,l^{\prime}}) \mathcal{CN}(0\mid \hat{r}_{i,j}^{w^{l,u,l^{\prime}}}, \tau_{i,j}^{w^{l,u,l^{\prime}}})
			+\zeta_{i,j}^{l,u,l^{\prime}}\rho_{i,j}^{\text{A}^{l,u,l^{\prime}}}}\nonumber \\
		&\bar{\omega}_{k,m,i,j}^{l,u,l^{\prime}} = 
		\frac{\omega_k^u \overleftarrow{p}^{l,u,l^{\prime}}_{m,i,j}
			\mathcal{CN}\left(\hat{r}_{i,j}^{w^{l,u,l^{\prime}}}\mid \mu^u_k a_m, \tau_{i,j}^{w^{l,u,l^{\prime}}}+\eta^u_k \left|a_m\right|^2 \right)}{\rho_{i,j}^{\text{A}^{l,u,l^{\prime}}}}\nonumber \\
		&\bar{\varphi}_{k,m,i,j}^{l,u,l^{\prime}}=
		(\frac{1}{\eta_k^u \left|a_m\right|^2} + \frac{1}{\tau_{i,j}^{w^{l,u,l^{\prime}}}})^{-1} \nonumber \\
		&\bar{\theta}_{k,m,i,j}^{l,u,l^{\prime}}=
		\bar{\varphi}_{k,m,i,j}^{l,u,l^{\prime}}
		(\frac{\mu_k^u}{\eta_k^u a_m^*} +
		\frac{\hat{r}_{i,j}^{w^{l,u,l^{\prime}}}}{\tau_{i,j}^{w^{l,u,l^{\prime}}}}).
	\end{align}
	The mean and variance with respect to (\ref{postw}) are given as
	\begin{align}
		\label{meanw}
		&\hat{w}_{i,j}^{l,u,l^{\prime}}=\chi_{i,j}^{l,u,l^{\prime}} \sum_{k=1}^{K}\sum_{m=1}^{\mathcal |A|}\bar{\omega}_{k,m,i,j}^{l,u,l^{\prime}}
		\bar{\theta}_{k,m,i,j}^{l,u,l^{\prime}}, \\
		&\hat{\tau}_{i,j}^{w^{l,u,l^{\prime}}}=\bar{\tau}_{i,j}^{w^{l,u,l^{\prime}}} -\left|\hat{w}_{i,j}^{l,u,l^{\prime}}\right|^2, \label{varw}
	\end{align}
	where $\bar{\tau}_{i,j}^{w^{l,u,l^{\prime}}}=\chi_{i,j}^{l,u,l^{\prime}}\sum_{k=1}^{K}\sum_{m=1}^{\mathcal |A|}\bar{\omega}_{k,m,i,j}^{l,u,l^{\prime}}(\left|\bar{\theta}_{k,m,i,j}^{l,u,l^{\prime}}\right|^2+\bar{\varphi}_{k,m,i,j}^{l,u,l^{\prime}})$.
	Similarly, the posterior distribution of $h_{i,j}^{l,u,l^{\prime}}$ is also approximated as a BGM distribution, given by
	\begin{align}
		\label{posth}
		&\Delta_{h_{i,j}^{l,u,l^{\prime}}}
		\propto \Delta^{h_{i,j}^{l,u,l^{\prime}}}_{g_{i,j}^{l,u,l^{\prime}}}
		\Delta_{h_{i,j}^{l,u,l^{\prime}}}^{g_{i,j}^{l,u,l^{\prime}}}
		= (1-\chi_{i,j}^{l,u,l^{\prime}})\delta(h_{i,j}^{l,u,l^{\prime}})
		+ \chi_{i,j}^{l,u,l^{\prime}} \times \nonumber \\ 
		&\qquad\sum_{k=1}^{K}\sum_{m=1}^{\mathcal |A|}\bar{\omega}_{k,m,i,j}^{l,u,l^{\prime}}\mathcal{CN}(w_{i,j}^{l,u,l^{\prime}}\mid \theta_{k,m,i,j}^{l,u,l^{\prime}}, \varphi_{k,m,i,j}^{l,u,l^{\prime}}),
	\end{align}
	where 
	\begin{align}
		&\varphi_{k,m,i,j}^{l,u,l^{\prime}}=
		(\frac{1}{\eta_k^u} + \frac{ \left|a_m\right|^2}{\tau_{i,j}^{w^{l,u,l^{\prime}}}})^{-1},\\
		&\theta_{k,m,i,j}^{l,u,l^{\prime}}=
		\varphi_{k,m,i,j}^{l,u,l^{\prime}}
		(\frac{\mu_k^u}{\eta_k^u} +
		\frac{\hat{r}_{i,j}^{w^{l,u,l^{\prime}}} a_m^*}{\tau_{i,j}^{w^{l,u,l^{\prime}}}}).
	\end{align}
	Then, the mean and variance with respect to (\ref{posth}) are given as 
	\begin{align}
		\label{channelE}
		&\hat{h}_{i,j}^{l,u,l^{\prime}}=\chi_{i,j}^{l,u,l^{\prime}} \sum_{k=1}^{K}\sum_{m=1}^{\mathcal |A|}\bar{\omega}_{k,m,i,j}^{l,u,l^{\prime}}
		\theta_{k,m,i,j}^{l,u,l^{\prime}},\\
		&\hat{\tau}_{i,j}^{h^{l,u,l^{\prime}}}=\bar{\tau}_{i,j}^{h^{l,u,l^{\prime}}} -\left|\hat{h}_{i,j}^{l,u,l^{\prime}}\right|^2,
	\end{align}
	where $\bar{\tau}_{i,j}^{h^{l,u,l^{\prime}}}=\chi_{i,j}^{l,u,l^{\prime}}\sum_{k=1}^{K}\sum_{m=1}^{\mathcal |A|}\bar{\omega}_{k,m,i,j}^{l,u,l^{\prime}}(\left|{\theta}_{k,m,i,j}^{l,u,l^{\prime}}\right|^2+{\varphi}_{k,m,i,j}^{l,u,l^{\prime}})$.
	After the algorithm convergence, the estimation of $h_{i,j}^{l,u,l^{\prime}}$ is given by (\ref{channelE}). In addition, the approximated posterior distribution of data symbols is given by
	\begin{align}
		\label{postt}
		\Delta_{t_{(l-l^{\prime})_M+uM}} = 
		\sum_{m=1}^{|\mathcal{A}|} 
		p^{l,u,l^{\prime}}_{m} \delta(t_{(l-l^{\prime})_M+uM}-a_m)	
	\end{align}
	where
	\begin{align}
		p^{l,u,l^{\prime}}_{m} \propto
		\overrightarrow{p}^{l,u,l^{\prime}}_{m,i,j}	\overleftarrow{p}^{l,u,l^{\prime}}_{m,i,j}.
	\end{align}
	Based on (\ref{postt}), we perform the symbol-by-symbol MAP estimation for data symbols as
	\begin{align}
		\label{tE}
		\hat{t}_{(l-l^{\prime})_M+uM}=\arg \max \limits_{a \in \mathcal A} \Delta_{t_{(l-l^{\prime})_M+uM}}(a).
	\end{align}
	\subsection{Learning Hyperparameters}
	EM algorithm is an iterative technique that increases a lower
	bound on the likelihood at each iteration, thus guaranteeing
	that the likelihood converges to a local maximum or at
	least a saddle point \cite{c25}. Based on the previous approximated posterior distributions, the EM algorithm is adopted to learn the phase rotation $\bm \phi \triangleq \{\bm\phi_q, \forall q\}$ and parameter $\bar{\mathbf p}\triangleq\{\sigma^2, \omega_k^u, \mu^u_k, \eta^u_k, \forall u,k\}$. Firstly, we assign the i.i.d. Gaussian distribution prior for the elements of $\bm \phi_q$, i.e., $\phi_{u,q}[l^{\prime}]\sim\mathcal{CN}(0,1)$, and treat $\mathbf W$ as the hidden variable. Then, the update of $\bm \phi_q$ in the $(t+1)$-th iteration is given by
	\begin{align}
		\label{EMphi}
		\bm \phi^{t+1}_q &= \arg \max \limits_{\bm \phi_q} 
		\mathrm{E}_{\mathbf W}[\log p(\mathbf W, \mathbf Y, \bm \phi_q) \mid \mathbf Y, \bm \phi^t] \\
		&= \arg \min \limits_{\bm \phi_q} \bm \phi_q^{\text H}\mathbf A_q\bm \phi_q - 2\Re\{\mathbf b_q^{\text H}\bm\phi_q\}\label{EMphi2}
	\end{align}
	where 
	\begin{align}
		&\mathbf A_q = \mathbf D^{\text T}[\sum_l (\mathbf C_q^l)^{\text H}\mathbf C_q^l \odot (\mathbf V_{\mathbf W}^l)^{\text T}]\mathbf D + \sigma^2 \mathbf I_{UM}, \\
		&\mathbf b_q = \mathbf D^{\text T}\mathrm{vec}_{\text d}(\sum_l (\mathbf C_q^l)^{\text H}\mathbf Y_q^l (\hat{\mathbf W}^l)^{\text H}),\\
		&\mathbf V_{\mathbf W}^l=\mathrm{diag}(\sum_j \mathrm{Var}[\mathbf w_j^l]) + \hat{\mathbf W}^l(\hat{\mathbf W}^l)^{\text H},\\
		&\mathbf D = \mathbf I_{UM}  \otimes \mathbf 1_N, 
	\end{align}
	$\mathbf w_j^l$ is the $j$-th column of $\mathbf W^l$, $\mathrm{Var}[\mathbf w_j^l]$ is the posterior variance of $\mathbf w_j^l$ given in (\ref{varw}), and $\mathbf 1_N \in \mathbb{R}^N$ is the all-one vector. 
	In addition, the optimization problem (\ref{EMphi2}) is subject to $\left|\phi_{u,q}[l^{\prime}]\right|=1$. Although the gradient-based algorithm can achieve a stationary point \cite{o7}, it involves large computations. Alternatively, we propose to adopt the following analytical rule, i.e.,
	\begin{align}
		\label{EMphi3}
		\bm \phi^{t+1}_q = \mathrm{Proj}(\mathbf A_q^{-1}\mathbf b_q),
	\end{align}
	where the $n$-th element of $\mathrm{Proj}(\mathbf x)$ is $x_n/\left|x_n\right|$. The motivation behind (\ref{EMphi3}) is that the optimal solution of (\ref{EMphi2}), when not constrained, is projected onto the unit circle in the complex plane, ensuring that $\left|\phi_{u,q}[l^{\prime}]\right|=1$. However, the matrix inversion required by (\ref{EMphi3}) typically entails substantial computational efforts, with a complexity of $\mathcal{O}(U^3M^3)$. Fortunately, as observed in (\ref{phiq}), $\phi_{u,q}[l^{\prime}]$ is nonzero only in the presence of a physical path corresponding to a delay of approximately $\frac{l^{\prime}}{M\Delta f}$. This sparsity can be exploited to truncate $\mathbf A_q$ and $\mathbf b_q$, thereby reducing the complexity involved in updating $\bm \phi_q$.
	In practical scenarios, the exact number of active physical paths $p_{\lambda}UP$, i.e., the number of nonzero elements in $\bm \phi_q$ is typically unknown. To address this, we retain $\bar P$ ($\bar P \geq p_{\lambda}UP$) elements of $\bm \phi_q$, selecting those corresponding to the indices of delay taps with the highest energies in $\mathbf W$. Initially, we calculate the energy of each delay tap as $\bm\iota=\{\iota^{l^{\prime}}_u\mid \iota^{l^{\prime}}_u=\sum_{l=0}^{M-1} \Vert\hat{\mathbf W}^{l,u,l^{\prime}}\Vert_{\text F}^2,\forall u,l^{\prime}\}$. We then form the set $\mathcal{E}$ containing the indices of the $\bar P$ largest elements in $\bm\iota$. Using $\mathcal{E}$, we select elements from $\mathbf A_q$ and $\mathbf b_q$ to construct $\bar{\mathbf A}_q \in \mathbb{C}^{\bar P \times \bar P}$ and $\bar{\mathbf b}_q \in \mathbb{C}^{\bar P}$, respectively. We denote the selected sub-vector of $\bm \phi_q$ according to $\mathcal{E}$ as $\bar{\bm \phi}_q$, which can be updated as
	\begin{align}
		\label{phiqupdate}
		\bar{\bm \phi}^{t+1}_q = \mathrm{Proj}(\bar{\mathbf A}_q^{-1}\bar{\mathbf b}_q).
	\end{align}
	Besides, the parameter $\bar{\mathbf p}$ is updated in the $(t+1)$-th iteration as
	\begin{align}
		\label{EMhyper}
		\bar{\mathbf p}^{t+1} = \arg \max \limits_{\bar{\mathbf p}} 
		\mathrm{E}_{\mathbf R, \mathbf H}[\log p(\mathbf R,\mathbf H, \mathbf Y \mid  \bar{\mathbf p}) \mid \mathbf Y, \bar{\mathbf p}^t],
	\end{align}
	where $\mathbf R = [\mathbf R^0, \dots, \mathbf R^{M-1}]$. 
	By examining the first derivative of the objective function with respect to the variables, the updates of parameters can be derived as
	\begin{align}
		\label{pstart}
		&(\sigma^2)^{t+1} = \frac{1}{QMNN_a} \sum_{l,o,j}  \left|y_{o,j}^l - \hat{r}^l_{o,j}\right|^2 + \tau^{r^l}_{o,j}\\
		&(\mu_k^u)^{t+1} = \frac{\sum_{l,l^{\prime},i,j}
			\chi_{i,j}^{l,u,l^{\prime}} \sum_{m=0}^{|\mathcal A|} \bar{\omega}_{k,m,i,j}^{l,u,l^{\prime}}
			\theta_{k,m,i,j}^{l,u,l^{\prime}}}{\sum_{l,l^{\prime},i,j} \chi_{i,j}^{l,u,l^{\prime}} \sum_{m=0}^{|\mathcal A|} \bar{\omega}_{k,m,i,j}^{l,u,l^{\prime}}
		}\\
		&(\eta_k^u)^{t+1} = \nonumber \\ 
		&\frac{\sum_{l,l^{\prime},i,j} \chi_{i,j}^{l,u,l^{\prime}} \sum_{m=0}^{|\mathcal A|} \bar{\omega}_{k,m,i,j}^{l,u,l^{\prime}}\left(
			\left|\theta_{k,m,i,j}^{l,u,l^{\prime}}-\mu_k^{u^t}\right|^2 + \varphi_{k,m,i,j}^{l,u,l^{\prime}} \right)}{\sum_{l,l^{\prime},i,j} \chi_{i,j}^{l,u,l^{\prime}} \sum_{m=0}^{|\mathcal A|} \bar{\omega}_{k,m,i,j}^{l,u,l^{\prime}}
		}\\
		&({\omega}_{k}^{u})^{t+1} = \frac{\sum_{l,l^{\prime},i,j} \chi_{i,j}^{l,u,l^{\prime}} \sum_{m=0}^{|\mathcal A|} \bar{\omega}_{k,m,i,j}^{l,u,l^{\prime}}}{\sum_{l,l^{\prime},i,j} \chi_{i,j}^{l,u,l^{\prime}}}\label{pend}
	\end{align}
	where $ \hat{r}^l_{o,j}$ and $\tau^{r^l}_{o,j}$ are updated by GAMP (see line 5 and 6 of Algorithm \ref{MRF-MP-GAMP}). 
	
	Building upon the message expressions and EM update
	rules, we propose the MRF-MP-GAMP algorithm for JDICESD as summarized in Algorithm \ref{MRF-MP-GAMP}, where we denote the index $uMN+l^{\prime}N+i$ as $v$. The lines 3-10 represent the GAMP algorithm, which can be executed in parallel for each $l$ and $j$. lines 11-18 are expressions derived using MP rules, and the lines 20 and 21 correspond to the EM update. Proper initialization of the
	hyperparameters is crucial for EM update which refers to \cite{c25}, and the damping is leveraged in the GAMP part to help the convergence of the algorithm \cite{g7}. 
	
	The computational complexity of the proposed algorithm is analyzed as follows: The computational complexity of the proposed receiver structure is primarily determined by the operations in the DDAI, JCESD, TSE modules, and the EM update. Specifically, the dominant computational task in the DDAI module is the matrix multiplication involved in the GAMP algorithm. For each iteration, this involves operations of complexity $\mathcal{O}(QUM^2N^2N_a)$. In the JCESD and TSE module, the complexity is governed by the element-wise product operations, which require $\mathcal{O}(\left|\mathcal{A}\right|QM^2NN_a)$ and $\mathcal{O}(UMNN_a)$ per iteration, respectively. For the EM update, the matrix inversion in (54) of our manuscript dominates the computations, with a complexity $\mathcal{O}(\bar P^3)$. Theoretically, $\bar P$ can be set to the number of active paths, i.e., $\bar P=p_{\lambda}UP$. In this case, the number of active paths is sufficiently small, e.g., $p_{\lambda}UP=12$, following the simulation parameters of our manuscript. As a result, the computational burden for the matrix inversion is negligible. In practical implementations, the exact number of active paths may be unknown. To handle this uncertainty, we recommend selecting a sufficiently large $\bar P$ to ensure that $\bar P \geq p_{\lambda}UP$. However, given the sparsity of active devices and physical paths in satellite channels, this should not pose a significant computational challenge. Overall, the total complexity of the MRF-MP-GAMP algorithm is in the order of $\mathcal{O}(QUM^2N^2N_a+\left|\mathcal{A}\right|QM^2NN_a+\bar P^3)$.
		
	Despite the iterative nature of the algorithm, these operations are well-suited for parallelization, as many tasks can be processed independently for each device or resource block. The parallelizable nature of the matrix multiplication and element-wise products in the DDAI, JCESD, and TSE modules allows for significant acceleration when implemented on graphics processing units (GPUs). Additionally, GPUs are also especially well-suited for the highly parallel structure of the DDAI module, where the GAMP algorithm can estimate $\mathbf W^l$ in parallel for each $l$. In particular, matrix operations can be distributed across multiple cores to achieve faster computation times. Addtionally, field-programmable gate arrays (FPGAs) can also be employed to further accelerate the proposed algorithms by implementing customized data paths for matrix operations and element-wise computations, enabling real-time processing with low latency. 
	
	\begin{algorithm}[hbt!]
		\footnotesize
		\caption{MRF-MP-GAMP}\label{MRF-MP-GAMP}
		\begin{algorithmic}[1]
			\REQUIRE{Received signals $\mathbf Y$, spreading code matrix $\mathbf C^l$, $\forall l$; the maximum number of iterations $\mathcal{I}_{\text out}$ and $\mathcal{I}_{\text{mrf}}$, and the termination threshold $\mathcal{T}_{\text out}$}
			\ENSURE The estimated activity indicator $\hat{\bm{\lambda}}$, effective channel $\hat{\mathbf H}$, and data symbols $\hat{\mathbf t}$.
			\STATE \textbf{Initialization}: $\forall u,l^{\prime},i,j$: Initialize $\xi^{\text{L}^{u,l^{\prime}}}_{i,j}=\xi^{\text{R}^{u,l^{\prime}}}_{i,j}=\xi^{\text{T}^{u,l^{\prime}}}_{i,j}=\xi^{\text{B}^{u,l^{\prime}}}_{i,j}=0.5$; $\forall l,o,j$: Initialize $\hat g^{l}_{o,j}=0$; $\forall l$: Initialize $\mathbf C^{\phi,l}=\mathbf C^l$; $\forall l,u,l^{\prime},i,j,m$: $\overleftarrow{p}^{l,u,l^{\prime}}_{m,i,j}=1/|\mathcal{A}|$; $\forall l,v,j$: choose $\hat{\tau}_{v,j}^{w^{l}}$, $\hat w^{l}_{v,j}$, and $\bar{\mathbf p}$. 
			\FOR{$t_{\mathrm{I}}=1$ to $\mathcal{I}_{\text out}$}
			\STATE $\forall l,o,j$: $\tau_{o,j}^{p^l}=\sum_{v}\left|c_{o,v}^{\phi,l}\right|^2 \hat{\tau}_{v,j}^{w^{l}}$
			\STATE $\forall l,o,j$: $\hat{p}_{o,j}^{l}=\sum_{v} c_{o,v}^{\phi,l} \hat w^{l}_{v,j}-\tau_{o,j}^{p^l} \hat g^{l}_{o,j}$.
			\STATE $\forall l,o,j,$: $\tau_{o,j}^{r^{l}}=\tau_{o,j}^{p^{l}}\sigma^2 / (\tau_{o,j}^{p^{l}}+\sigma^2)$ 
			\STATE $\forall l,o,j,$: $\hat r_{o,j}^{l}=(\tau_{o,j}^{p^{l}}y_{o, j}^{l}+\sigma^2 \hat{p}_{o,j}^{l})/(\tau_{o,j}^{p^{l}}+\sigma^2)
			$ 
			\STATE $\forall l,o,j, $: $\tau_{o,j}^{g^{l}}=(\tau_{o,j}^{p^{l}}-\tau_{o,j}^{r^{l}}) /(\tau_{o,j}^{p^{l}})^2$
			\STATE $\forall l,o,j, $: $\hat g^{l}_{o,j}=(\hat r_{o,j}^{l}-\hat{p}_{o,j}^{l}) / \tau_{o,j}^{p^{l}}$
			\STATE $\forall l,v,j$: $\tau^{w^{l}}_{v,j}=(\sum_o\left|c_{o,v}^{\phi,l}\right|^2 \tau_{o,j}^{g^{l}})^{-1}$
			\STATE $\forall l,v,j$: $\hat r^{w^{l}}_{v,j}=\hat w^{l}_{v,j}+\tau^{w^{l}}_{v,j} \sum_o c_{o,v}^{\phi,l^*} \hat g^{l}_{o,j}$
			\STATE $\forall l,u,l^{\prime},i,j$: Compute $\rho_{i,j}^{l,u,l^{\prime}}$ via (\ref{rho})
			\FOR{$t_{\text{mrf}}=1$ to $\mathcal{I}_{\text{mrf}}$}
			\STATE $\forall u,l^{\prime},i,j$: Update $\xi^{\text{L}^{u,l^{\prime}}}_{i,j},\xi^{\text{R}^{u,l^{\prime}}}_{i,j},\xi^{\text{T}^{u,l^{\prime}}}_{i,j},\xi^{\text{B}^{u,l^{\prime}}}_{i,j}$ via (\ref{xiupdate})
			\ENDFOR
			\STATE $\forall l,u,l^{\prime},i,j,$: Update $\zeta_{i,j}^{l,u,l^{\prime}}$ via (\ref{zitaupdate}) 
			\STATE $\forall l,u,l^{\prime},i,j,m$: Update$\overrightarrow{p}^{l,u,l^{\prime}}_{m,i,j}$ and $\overleftarrow{p}^{l,u,l^{\prime}}_{m,i,j}$ via (\ref{pright}) and (\ref{pleft})
			\STATE $\forall l,u,l^{\prime},i,j,m$: Compute $\chi_{i,j}^{l,u,l^{\prime}}$, $\bar{\omega}_{k,m,i,j}^{l,u,l^{\prime}}$,
			$\bar{\varphi}_{k,m,i,j}^{l,u,l^{\prime}}$, and
			$\bar{\theta}_{k,m,i,j}^{l,u,l^{\prime}}$ via (\ref{poststart})
			\STATE $\forall l,u,l^{\prime},i,j$: Update $\hat{w}_{i,j}^{l,u,l^{\prime}}$,  $\hat{\tau}_{i,j}^{w^{l,u,l^{\prime}}}$, and $\hat{h}_{i,j}^{l,u,l^{\prime}}(t_{\mathrm{I}}+1)$ via (\ref{meanw}), (\ref{varw}), and (\ref{channelE})
			\STATE \textbf{if} $\sum_l\left\|\hat{\mathbf H}^l(t_{\mathrm{I}}+1)-\hat{\mathbf H}^l(t_{\mathrm{I}})\right\|^2_{\mathrm{F}} \leq \mathcal{T}_{\text out} \sum_l\left\|\hat{\mathbf H}^l(t_{\mathrm{I}})\right\|^2_{\mathrm{F}}$: break \textbf{end if}
			\STATE $\forall q$: Update $\bar{\bm \phi}_q$ via (\ref{phiqupdate})
			\STATE Update $\bar{\mathbf p}$ via (\ref{pstart})-(\ref{pend})
			\STATE $\forall l$: Reconstruct $\mathbf C^{\phi,l}$ via (\ref{Qframe})
			\ENDFOR
			\STATE $\forall u$: Compute $\hat \lambda_u$ via (\ref{detectA})
			\STATE $\forall l,l^{\prime},u$: Compute $\hat{t}_{(l-l^{\prime})_M+uM}$ via (\ref{tE})
		\end{algorithmic}
	\end{algorithm}
	
	\section{CNN-Enhanced Detector}
	\label{CNN-Enhanced Detector}
	CNNs are widely being used in many DL applications due to their remarkable ability to combine feature extraction and detection tasks. In this section, we adopt CNNs to efficiently exploit the statistical information provided by the MRF-MP-GAMP algorithm, enhancing symbol detection performance.
	
	\subsection{CNN Framework}
	\subsubsection{Preprocessing}
	We denote the estimated set of active devices as $\hat{\mathcal{U}}=\{u|\hat{\lambda}_u=1,u=0,\cdots,U-1\}$. According to the equality constraint in (\ref{auxiliary}), for each $u \in \mathcal{U}$, a simple data detection rule can be represented as $\hat t_{(l-l^{\prime})_M+uM}=w^{l,u,l^{\prime}}_{i,j}/h_{i, j}^{l, u, l^{\prime}}$ given the perfect $w^{l,u,l^{\prime}}_{i,j}$ and $h_{i, j}^{l, u, l^{\prime}}$. However, the perfect channel information is hard to get directly in the practical scenarios. Fortunately, the proposed MRF-MP-GAMP algorithm can output the approximated MMSE estimations $\hat{w}_{i,j}^{l,u,l^{\prime}}$ and $\hat{h}_{i,j}^{l,u,l^{\prime}}$, and the corresponding estimation uncertainty $\hat{\tau}_{i,j}^{w^{l,u,l^{\prime}}}$ and $\hat{\tau}_{i,j}^{h^{l,u,l^{\prime}}}$. We expect that CNNs exploit this statistical information to learn an efficient detection rule. Specifically, we collect all the effective channel elements related to the data symbol $t_{l+uM}$ into the 3D tensor $\check{\mathbf{H}}^{l,u}\in \mathbb{C}^{N\times N_a \times M}$ with $\check{\mathbf{H}}^{l,u}[:,:,\hat l]=\hat{\mathbf{H}}^{\hat l,u,(\hat l - l)_M}$, $\hat l={0,\cdots,M-1}$. Similarly, we can get $\check{\mathbf{W}}^{l,u}$, $\check{\mathbf T}_w^{l,u}$, and $\check{\mathbf T}_h^{l,u}$, where  $\check{\mathbf{W}}^{l,u}[:,:,\hat l]=\hat{\mathbf{W}}^{\hat l,u,(\hat l - l)_M}$, $\check{\mathbf T}_w^{l,u}[:,:,\hat l]$ is with the $(i,j)$-th element $\hat{\tau}_{i,j}^{w^{\hat l,u,(\hat l - l)_M}}$, and $\check{\mathbf T}_h^{l,u}[:,:,\hat l]$ is with the $(i,j)$-th element $\hat{\tau}_{i,j}^{h^{\hat l,u,(\hat l - l)_M}}$. Then, we stack the four tensors along the last dimension to get the input of the CNN as $\check{\mathbf{I}}_c^{l,u}=[[\check{\mathbf{W}}^{l,u}],[\check{\mathbf T}_w^{l,u}],[\check{\mathbf{H}}^{l,u}],[\check{\mathbf T}_h^{l,u}]] \in \mathbb{C}^{N\times N_a \times 4M}$. Note that the statistical information for each transmitted symbols is extracted individually from the outputs of the Algorithm \ref{MRF-MP-GAMP}. Then, the CNN could efficiently estimate all transmitted symbols in parallel and accommodate the dynamic number of active devices.
	
	\subsubsection{Training phase}
	In this phase, by adjusting the weights of the adopted CNN, we aim to make the output of the CNN as close as possible to the true posterior distribution of the transmitted symbols. The dataset $\Theta=\bigcup\limits_n \Theta_n$ is firstly constructed, where $\Theta_n=\{(\check{\mathbf{I}}_{c_n}^{l,u}, {\mathbf p}_{t_n}^{l,u})| u\in \mathcal{U}_n\cap\hat{\mathcal{U}}_n, l=0,\dots,M-1\}$ is generated from the $n$-th sample $(\mathbf Y_n, \mathbf t_n)$. Here, $\mathbf Y_n$ and $\mathbf t_n=[\mathbf{t}_{n_0},\dots,\mathbf{t}_{n_{U-1}}]$ represent the received signal and the transmitted symbols of all devices, respectively. $\check{\mathbf{I}}_{c_n}^{l,u}$ is the input to the CNN, while ${\mathbf p}_{t_n}^{l,u}$ is the true posterior distribution of the transmitted symbols and serves as the target of CNN. In addition, $\mathcal{U}_n=\{u|\lambda_{n_u}=1,u=0,\dots,U-1\}$ is denoted as the set of true active devices. 
		
		To construct the sub-dataset $\Theta_n$, $\mathcal{U}_n$ is initially determined based on the device activity $p_{\lambda}$. Then, the transmitted symbols $\mathbf t_n$ are generated, with symbols for active devices randomly selected from the predefined alphabet $\mathcal{A}$ and those for inactive devices set to zero. Subsequently, received signal $\mathbf Y_n$ is obtained by transmitting $\mathbf t_n$ over a randomly time-varying TSL using the transmission scheme proposed in Sec. \ref{transmission scheme}. By feeding $\mathbf Y_n$ into Algorithm \ref{MRF-MP-GAMP}, the set of estimated active devices $\hat{\mathcal{U}}_n$, the MMSE estimations $\hat{w}_{n_{i,j}}^{l,u,l^{\prime}}$ and $\hat{h}_{n_{i,j}}^{l,u,l^{\prime}}$, and the corresponding estimation uncertainties $\hat{\tau}_{n_{i,j}}^{w^{l,u,l^{\prime}}}$ and $\hat{\tau}_{n_{i,j}}^{h^{l,u,l^{\prime}}}$ are obtained. Following the preprocessing module, $\check{\mathbf{I}}_{c_n}^{l,u}$ is derived for $u\in \mathcal{U}_n\cap\hat{\mathcal{U}}_n$ and $l=0,\dots,M-1$. Furthermore, the one-hot encoding is adopted to represent the true posterior distribution for the transmitted symbols, i.e., ${\mathbf p}_{t_n}^{l,u}=[\mathbb{I}\{t_{n_{l+uM}}=a_1\},\cdots,\mathbb{I}\{t_{n_{l+uM}}=a_{\left|\mathcal{A}\right|}\}]$ for $u\in \mathcal{U}_n\cap\hat{\mathcal{U}}_n$, and $l=0,\dots,M-1$, where $t_{n_{l+uM}}$ denotes the $(l+uM)$-th element of $\mathbf t_n$.
		
		The CNN, parameterized by weights $\bm \varrho$, defines the non-linear function $f(\cdot;\bm \varrho)$, which maps $\check{\mathbf{I}}_{c_n}^{l,u}$ to the estimated posterior distribution $\hat{{\mathbf p}}_{t_n}^{l,u}$ for the transmitted symbol $t_{n_{l+uM}}$, i.e.,
		\begin{align}
			\hat{{\mathbf p}}_{t_n}^{l,u} = f(\check{\mathbf{I}}_{c_n}^{l,u};\bm \varrho).
		\end{align}
		To optimize the weights of CNN, the cross entropy is utilized as the loss function, i.e.,
		\begin{align}
			\label{loss}
			\bm \varrho^* = \arg \min \limits_{\bm \varrho} \mathrm{E}[-\sum_m {p}_{t_{n_m}}^{l,u} \log \hat{{p}}_{t_{n_m}}^{l,u}],
		\end{align}
		where ${p}_{t_{n_m}}^{l,u}$ and $\hat{{p}}_{t_{n_m}}^{l,u}$ are the $m$-th element of ${\mathbf p}_{t_n}^{l,u}$ and $\hat{{\mathbf p}}_{t_n}^{l,u}$, respectively, and the expectation is calculated over $\Theta$. In the simulations, the weights are updated iteratively using the Adam optimizer \cite{o11}. During each iteration, only a small batch of elements from $\Theta$ are adopted to calculate the expectation in (\ref{loss}).
	
	\subsubsection{Testing phase}
	In this phase, the received signal $\mathbf Y$ is firstly processed by Algorithm \ref{MRF-MP-GAMP}, yielding outputs $\hat{\mathcal{U}}$, $\hat{w}_{i,j}^{l,u,l^{\prime}}$, $\hat{h}_{i,j}^{l,u,l^{\prime}}$, $\hat{\tau}_{i,j}^{w^{l,u,l^{\prime}}}$, and $\hat{\tau}_{i,j}^{h^{l,u,l^{\prime}}}$. Subsequently, $\check{\mathbf{I}}_{c}^{l,u}$ is constructed for $u \in \hat{\mathcal{U}}$ and $l=0,\dots,M-1$, which is then fed into the trained CNN to get the estimated posterior distribution $\hat{{\mathbf p}}_{t}^{l,u}$ for the transmitted symbol $t_{l+uM}$. Finally, the detection of symbol $t_{l+uM}$ is performed by choosing $\hat t_{l+uM} = a_{m^*}$ with 
	\begin{align}
		\label{tRefine}
		m^*=\arg \max \limits_{m} \hat{{p}}_{t_m}^{l,u}.
	\end{align}
	Further, the estimated effective channel is refined by the more precise data symbols, as given by
	\begin{align}
		\label{hrefine}
		\hat{h}_{i,j}^{\hat l,u,(\hat l - l)_M} = \hat{w}_{i,j}^{\hat l,u,(\hat l - l)_M} / \hat t_{l+uM}.
	\end{align}

	\begin{figure}[!htb]
		\centering
		\captionsetup{font={small}}
		\includegraphics[width=3.45in]{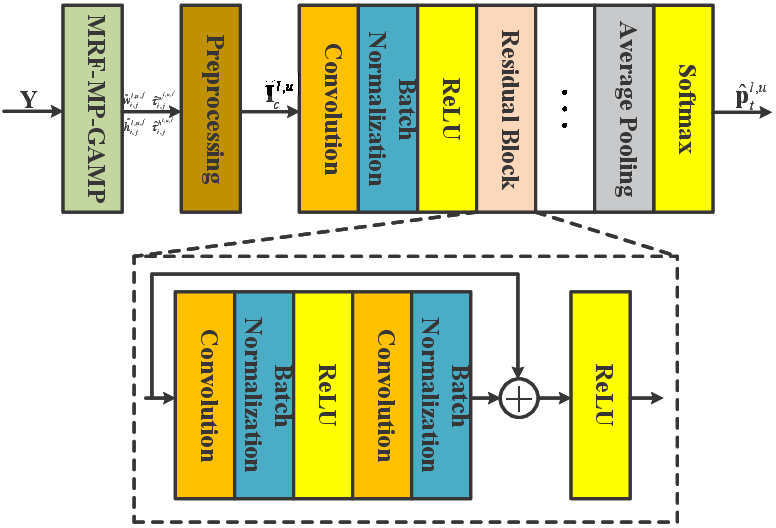}
		\caption{The architecture of MRF-MP-GAMP-CNN.}
		\label{MRF-MP-GAMP-CNN}
	\end{figure}
	\subsection{Architecture of MRF-MP-GAMP-CNN}
	As illustrated in Fig. \ref{MRF-MP-GAMP-CNN}, the proposed MRF-MP-GAMP-CNN framework comprises three main modules. Initially, the MRF-MP-GAMP module produces the set of estimated active devices, and the posterior means and variances for $\mathbf{W}$ and $\mathbf{H}$ of active devices. Subsequently, these outputs are combined in the preprocessing module to form the input for the CNN module. This module leveraging statistical information to improve symbol detection, involves a convolutional layer, batch normalization, and the ReLU activation function as its foundational elements. Additionally, residual learning \cite{o8} is incorporated to facilitate the optimization for CNN weights. The output layer employs a Softmax function to estimate the posterior distribution of the data symbols. Finally, the refined data symbols and effective channel are represented by (\ref{tRefine}) and (\ref{hrefine}), respectively. It is worth noting that this model utilizes a complex neural network \cite{o9}, which defines the operations in the complex plane, such as complex convolution. Unlike traditional methods, this approach allows the algorithm to process complex numbers directly, eliminating the need to convert input signals into the real domain.

		\subsection{Practical Deployment Considerations}
		
		\subsubsection{Impact on satellite payload}
		The proposed receiver structure, despite its iterative nature, can be accelerated using modern hardware like GPUs and FPGAs, which offer high computational performance with minimal physical footprint. Advances in miniaturization allow powerful units to be integrated into satellites without significantly increasing payload mass. Furthermore, the hardware can be scaled according to mission requirements, with fewer processing units for lower communication demands and more for higher demands.
		
		\subsubsection{Power consumption}
		Power consumption is critical for LEO satellites with limited onboard resources. While GPUs and FPGAs offer computational advantages, they also introduce power considerations. However, modern versions are energy-efficient, and power management techniques like dynamic voltage and frequency scaling (DVFS) \cite{g5} can reduce power usage during low-demand periods. Low-power FPGAs designed for space applications could further minimize power consumption. Another factor mitigating power concerns is that, in practice, the data from active devices may be sent over $kQ$ frames, where $k\in \mathbb{N}$. Once the device states (active or inactive) are identified after the first $Q$ frames, the remaining $(k-1)Q$ frames are only used for joint channel estimation and symbol detection for active devices. In this case, the received signal and spreading code of the active devices are input into the MRF-MP-GAMP-CNN, which then performs computations only for the active devices, rather than for all potential devices. This optimization ensures that the full computational power is not required throughout the entire transmission, helping to conserve energy.
		
		\subsubsection{Feasibility of integration with existing satellite infrastructure}
		As indicated in \cite{o2} and \cite{o3}, current 3GPP NTN protocols do not fully support IoT devices without GNSS, while this is a focus of ongoing standardization efforts. Fortunately, only minor changes to satellite infrastructure are needed, such as extending CP length to exceed maximum differential delay and adding ISFFT and SFFT modules to support OTFS-based transmissions. Regarding the integration of the proposed algorithm, especially given the ongoing evolution of satellite technology. Many modern LEO satellite platforms already support advanced onboard processing, including hardware-accelerated systems.
		
		Overall, while challenges related to payload, power consumption, and integration exist, they are manageable with current technology and design optimizations. The proposed system is scalable and suitable for future satellite deployments supporting remote IoT applications.

	\begin{table}
		\scriptsize
		\caption{{\sc Simulation Parameters}}
		\begin{center}
			\begin{tabular}{|c|c|}
				\hline Parameter & Values \\
				\hline Operating band & S-band \\
				\hline Power delay profile & NTN-TDL-D \\
				\hline Modulation Scheme & 4NPAM \\
				\hline Frame size $(M, N)$ & $(16, 7)$ \\
				\hline Subcarrier spacing $\Delta f$ & 30 kHz \\
				\hline Orbit altitude & 600 km \\
				\hline Spot beam radius & 128 km \\
				\hline Minimum elevation angle & 30° \\
				\hline Differential delay & $[0,699]$ \textmu s\\
				\hline Doppler shift & $[-41,41]$ kHz\\
				\hline Zenith angle & $[-\pi/2, \pi/2)$\\
				\hline Azimuth angle & $[0,2\pi)$\\
				\hline Rician factor  & 11.707 dB\\
				\hline
			\end{tabular}
			\label{spa}
		\end{center}
	\end{table}
	
	\begin{table}
	
		\footnotesize
		\centering 
		\caption{{\sc Basic Parameters of CNN}}
		\label{parameters of CNN} 
		\begin{tabular}{m{2.3cm}<{\centering}|m{2.2cm}<{\centering}|m{2.2cm}<{\centering}}
			\hline 
			Layer& Parameters of filter& Activation function\\ 
			\hline 
			Convolutional layer& (64, $3 \times 3$)& ReLU\\	
			\hline
			
			Residual block 1& (64, $3 \times 3$)& ReLU \\ 
			\hline
			
			Residual block 2& (64, $3 \times 3$)& ReLU \\ 
			\hline
			
			Residual block 3& (128, $3 \times 3$)& ReLU \\ 
			\hline
			
			Residual block 4& (128, $3 \times 3$)& ReLU \\ 
			\hline
			
			Residual block 5& (256, $3 \times 3$)& ReLU \\ 
			\hline
			
			Residual block 6& (256, $3 \times 3$)& ReLU \\ 
			\hline
			
			Residual block 7& (512, $3 \times 3$)& ReLU \\ 
			\hline
			
			Residual block 8& (512, $3 \times 3$)& ReLU \\ 
			\hline
			
			Average pooling& (512, $N \times N_yN_z$)&  \\ 
			\hline
			
			Output layer& ($|\mathcal{A}|$, )& Softmax \\ 
			\hline	
		\end{tabular} 
	\end{table}
	
	\section{Numerical Results}
	\label{Numerical Results}
	
	This section presents simulation results to demonstrate the effectiveness of the proposed algorithms. We consider a typical random access scenario in the non-terrestrial networks without GNSS, where $U=40$ potential devices are randomly distributed within the coverage area of the satellite. This area is defined as a circle with the radius of 128 km. The satellite is equipped with $4\times 4$ UPA and the device activity is set as $p_{\lambda}=0.1$ unless otherwise specified. The power delay profile of TSL follows the NTN-TDL-D\cite{3gpp}. The nonnegative pulse amplitude
	modulation (NPAM) is adopted to alleviate the phase ambiguity problem, which has been applied broadly in the non-coherent communications\cite{o10}. Since large-scale propagation effects vary over relatively long distances, we assume that these effects can be effectively compensated \cite{ss32}. The main parameters are summarized in Table \ref{spa}. In addition, the elements of each device's spreading code follow the i.i.d. Gaussian distributions, i.e., $C_{u,q}[k,l]\sim\mathcal{CN}(0,\frac{1}{QN})$. The received signal-to-noise ratio is defined as $\text{ SNR }=10\log_{10}\frac{\sum_{l=0}^{M-1}\Vert \mathbf R^l\Vert^2_F}{QMNN_a\sigma^2}$. For the MRF-MP-GAMP-CNN, datasets are generated with 8,000 samples for training, 2,000 for validation, and 2,000 for testing. The CNN architecture consists of 8 residual blocks and the Adam optimizer\cite{o11} is utilized to update the weights at a learning rate of 0.0001 and a batch size of 128. All the basic parameters of CNN are given in Table \ref{parameters of CNN}, where ($\mathcal{C}$, $\mathcal{W}\times \mathcal{W}$) indicates that the number of output channels is $\mathcal{C}$ and the filter size is $\mathcal{W}\times \mathcal{W}$.
	
	The activity error rate (AER), normalized mean-squared-error (NMSE), and symbol error rate (SER) are adopted as metrics for device identification, channel estimation, and symbol detection, respectively, given by $\text{AER} = \frac{1}{U}\sum_{u=0}^{U-1}\left|\lambda_u-\hat \lambda_u\right|$,
	$\text{NMSE}=\frac{\sum_{l=0}^{M-1}\left\| \mathbf H^l-\hat{\mathbf H}^l\right\|^2_{\mathrm{F}}}{\sum_{l=0}^{M-1}\left\|\mathbf H^l\right\|^2_{\mathrm{F}}}$, and
	$\text{SER} = \frac{1}{UM}\sum_{i=0}^{UM-1} \mathbb{I}\{t_i \neq \hat t_i\}$, where we assume inactive devices transmit zero for computing the SER. The two-phase OTFS scheme is adopted for performance comparison, incorporating three benchmarks. The first benchmark, ConvSBL-GAMP+MP, utilizes the ConvSBL-GAMP\cite{c11} for JDICE, followed by symbol detection using the MP algorithm \cite{o5}. The second, GMMV-AMP+GAMP, employs the GMMV-AMP\cite{s43} for similar purposes, with symbol detection conducted via the GAMP algorithm \cite{sss}. The third, Oracle-LMMSE, leverages perfect channel state information to detect data symbols using the LMMSE equalizer \cite{c22}. We evaluate the proposed algorithms in two distinct scenarios: (i) The on-grid case, where the fractional Doppler $\tilde{k}_{i,u}=0$, does not involve the extra phase rotation. (ii) The off-grid case, where $\tilde{k}_{i,u} \in (-\frac{1}{2},\frac{1}{2}]$, requires to update the phase rotation by the proposed algorithm. Note that conventional algorithms are not directly applicable to multi-frame transmission scenarios affected by fractional Doppler shifts, and thus they are evaluated solely in the on-grid case.
	
	\begin{figure}[!htb]
		\centering
		\captionsetup{font={small}}
		\setlength{\belowcaptionskip}{-.2cm}
		\includegraphics[width=3in]{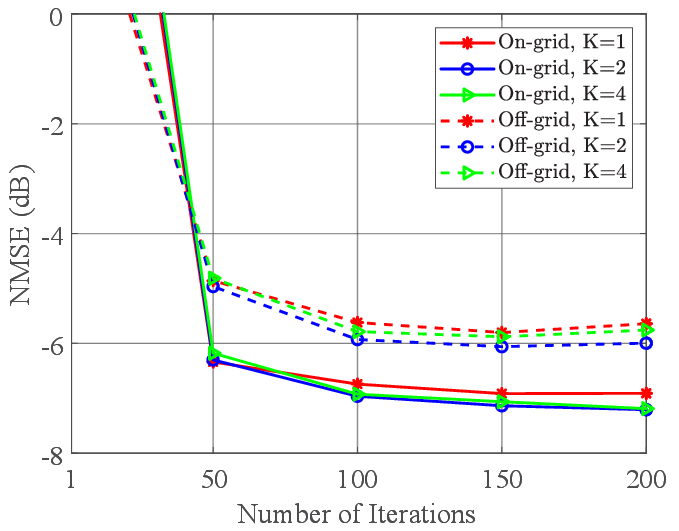}
		\caption{The convergence of MRF-MP-GAMP, where $\text{SNR}=2$ dB and $Q=8$.}
		\label{convergence}
	\end{figure}
	Fig. \ref{convergence} shows the convergence of the proposed MRF-MP-GAMP, which plots NMSE versus the number of iterations at $\text{SNR}=2$ dB and $Q=8$. This figure also illustrates the impact of varying the number of BGM components $K$ on the algorithms performance. It is observed that the proposed algorithms tends to be stable after approximately 100 iterations, and its convergence rate appears to be relatively insensitive to changes in $K$. Besides, the marginal gain can be achieved by increasing $K$ from 1 to 2. However, increasing $K$ to 4 results in performance degradation, likely due to overfitting effects associated with a larger $K$. Based on these observations, subsequent simulations are conducted with $K=2$ and the number of iterations $\mathcal{I}_{\text out}=150$.
	
	\begin{figure}
		\centering
		\captionsetup{font={small}}
		\subfigure[Channel estimation performance.]{
			\label{Frame_NMSE}
			\begin{minipage}{7.6cm}
				\includegraphics[width=\textwidth]{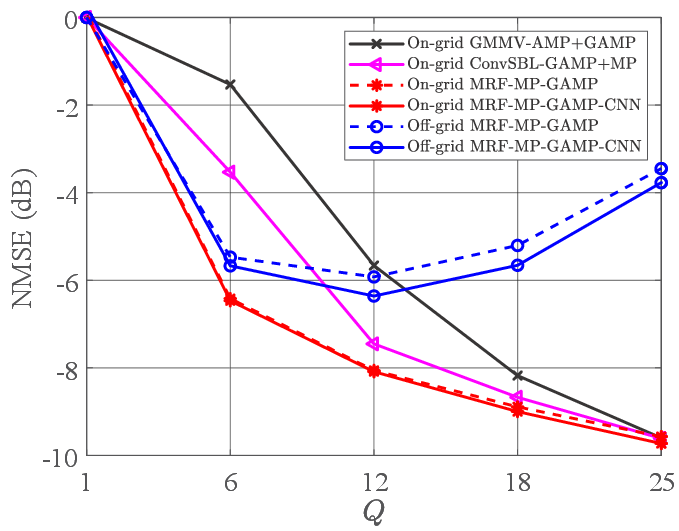} \\
				\vspace{-0.2cm}	
			\end{minipage}
		}
		\subfigure[Symbol detection performance.]{
			\label{Frame_SER}
			\begin{minipage}{7.6cm}
				\includegraphics[width=\textwidth]{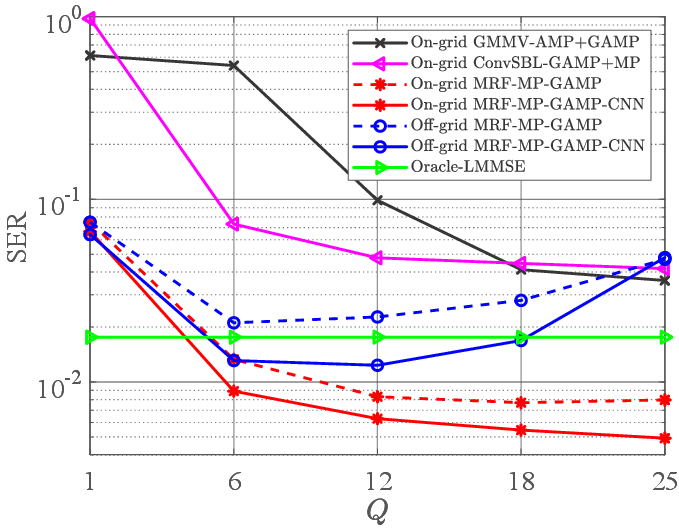} \\
				\vspace{-0.2cm}	
			\end{minipage}
		}
		\vspace{-0.2cm}
		\caption{Performance comparison among different schemes given different number of frames, where $\text{SNR}=2$ dB.} 
		\vspace{0.6cm}
		\label{FramePer}
	\end{figure}
	Fig. \ref{FramePer} illustrates the influence of the number of frames $Q$ on the performance of the proposed algorithms, presenting both NMSE and SER as functions of $Q$. Notably, the single-frame transmission scheme $(Q=1)$ performs inadequately in scenarios with large differential delay and Doppler shift.  In the on-grid case, as $Q$ increases, it is observed that the NMSE and SER of all the schemes continually decrease, and the proposed algorithms always perform much better than the benchmarks. This indicates the effectiveness the proposed multi-frame transmission scheme and the JDICESD algorithms. For example, at $Q=12$, the SER of the MRF-MP-GAMP is lower than 0.01 and is one-fifth that of the ConvSBL-GAMP+MP; the NMSE of the MRF-MP-GAMP is about 0.5 and 2 dB lower than the ConvSBL-GAMP+MP and GMMV-AMP+GAMP, respectively. Additionally, the MRF-MP-GAMP-CNN benefits from further exploiting statistical information, achieving additional gains over its counterparts. In the off-grid case, while increased $Q$ values initially reduce NMSE and SER for proposed algorithms approximately up to $Q=12$, further increases in $Q$ degrade performance due to the rising number of unknown phase rotations, which pose a significant estimation challenge. In practical communications, smaller $Q$ values are preferred to maintain the quasi-static properties of the channel in the delay-Doppler domain. Despite performance declines, the MRF-MP-GAMP-CNN still outperforms the Oracle-LMMSE in scenarios where $6\leq Q\leq 18$. In subsequent simulations, we will fix $Q=8$ to ensure performance and computational efficiency.
	
	\begin{figure}
		\centering
		\captionsetup{font={small}}
		\setlength{\belowcaptionskip}{-.6cm}
		\subfigure[Channel estimation performance.]{
			\label{SNR_NMSE}
			\begin{minipage}{7.6cm}
				\includegraphics[width=\textwidth]{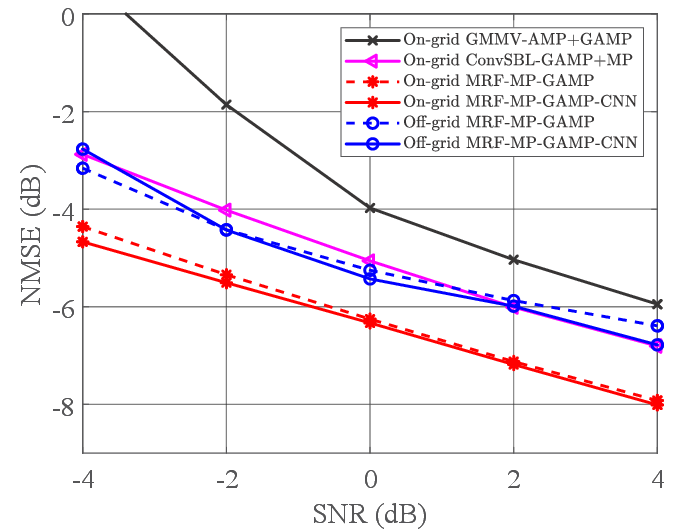} \\
				\vspace{-0.2cm}	
			\end{minipage}
		}
		\subfigure[Symbol detection performance.]{
			\label{SNR_SER}
			\begin{minipage}{7.6cm}
				\includegraphics[width=\textwidth]{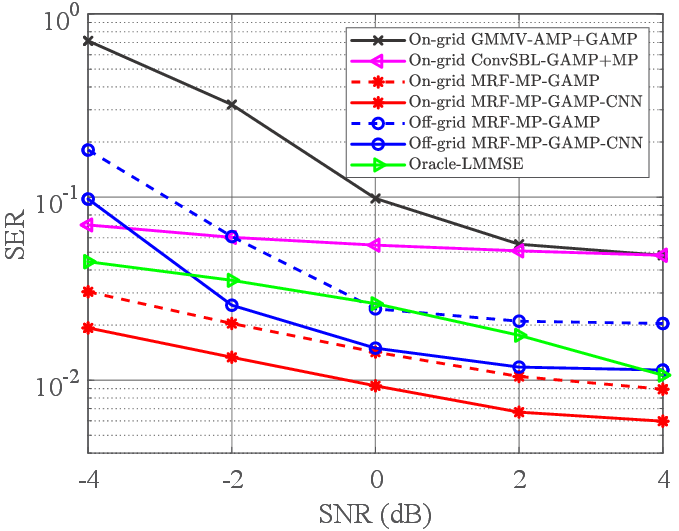} \\
				\vspace{-0.2cm}	
			\end{minipage}
		}
		\vspace{-0.2cm}
		\caption{Performance comparison among different schemes given different SNR values, where $Q=8$.} 
		\vspace{0.6cm}
		\label{SNRPer}
	\end{figure}
	Fig. \ref{SNRPer} depicts the performance of various algorithms in terms of NMSE and SER as functions of SNR. It is observed that the proposed algorithms significantly outperform the benchmarks in the on-grid case. Specifically, Fig. \ref{SNR_NMSE} shows that at 0 dB SNR, the proposed MRF-MP-GAMP improves upon ConvSBL-GAMP+MP and GMMV-AMP+GAMP by approximately 1.2 dB and 2.2 dB in NMSE, respectively; Fig. \ref{SNR_SER} reveals that at an SER of 0.01, MRF-MP-GAMP surpasses Oracle-LMMSE by about 2 dB in terms of SNR.
	Conventional schemes that operate separately have high error floors for SER, limiting their ability to support a large number of devices. In contrast, the proposed algorithms, benefited by JDICESD design, exhibit robust performance. Furthermore, Fig. \ref{SNRPer} indicates that in the off-grid case, the NMSE for MRF-MP-GAMP is lower than that for on-grid GMMV-AMP+GAMP and comparable to on-grid ConvSBL-GAMP+MP. Additionally, the SER for off-grid MRF-MP-GAMP is lower than that of both GMMV-AMP+GAMP and ConvSBL-GAMP+MP when the SNR exceeds -2 dB, validating the effectiveness of the proposed scheme for LEO satellite-based uplink transmission in the presence of large differential delay and Doppler shift. Moreover, the SER for off-grid MRF-MP-GAMP-CNN outperforms Oracle-LMMSE at SNRs above -2 dB, which demonstrates the capability of the CNN-enhanced symbol detector.
	
	\begin{figure}
		\centering
		\captionsetup{font={small}}
		\setlength{\belowcaptionskip}{-.6cm}
		\subfigure[Channel estimation performance.]{
			\label{Activity_NMSE}
			\begin{minipage}{7.7cm}
				\includegraphics[width=\textwidth]{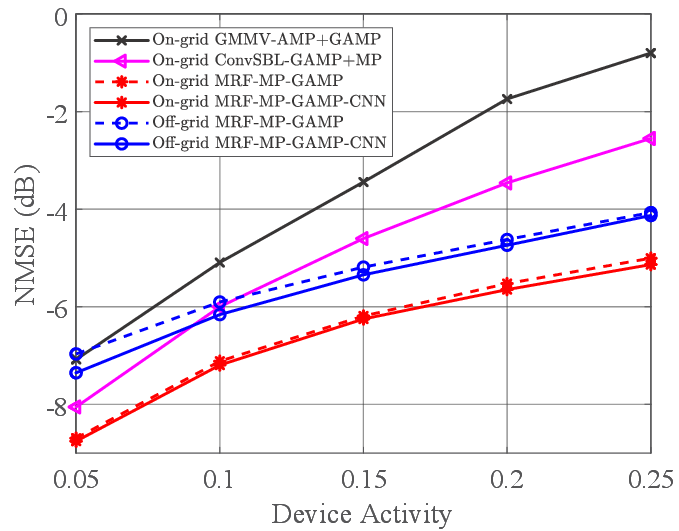} \\
				\vspace{-0.2cm}	
			\end{minipage}
		}
		\subfigure[Symbol detection performance.]{
			\label{Activity_SER}
			\begin{minipage}{7.7cm}
				\includegraphics[width=\textwidth]{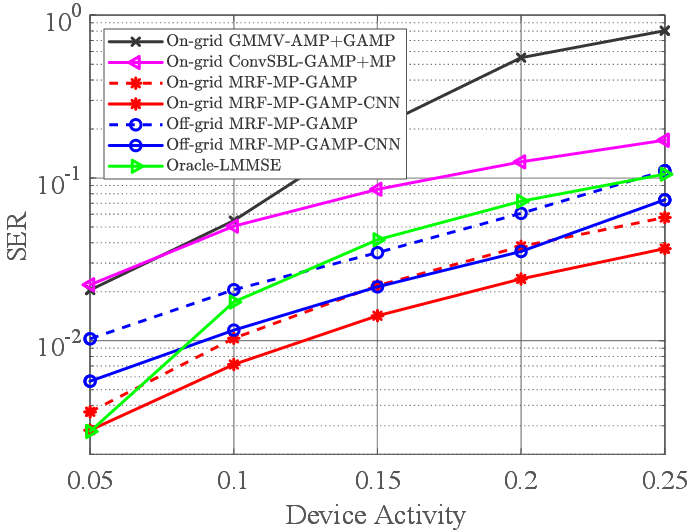} \\
				\vspace{-0.2cm}	
			\end{minipage}
		}
		\vspace{-0.2cm}
		\caption{Performance comparison among different schemes given different device activities, where $\text{SNR}=2$ dB and $Q=8$.} 
		\vspace{0.6cm}
		\label{ActivityPer}
	\end{figure}
	Fig. \ref{ActivityPer} plots the NMSE and SER as the functions of the device activity, given $\text{SNR}=2$ dB. It is observed that the on-grid MRF-MP-GAMP-CNN achieves the best performance. Notably, the off-grid MRF-MP-GAMP-CNN outperforms the three on-grid benchmarks when device activity exceeds 0.1, which indicates the proposed schemes are capable of supporting more active devices access. Note that the performance of all the schemes deteriorates obviously when the device activity increases. This degradation is primarily due to the limited size of the receiving antenna array, which results in decreased spatial separation as the number of active devices grows. Consequently, inter-user interference intensifies in the spatial domain, further impacting performance.
	
	\begin{figure}
		\centering
		\captionsetup{font={small}}
		\setlength{\belowcaptionskip}{-.6cm}
		\subfigure[Device identification performance under different SNR values, where $p_{\lambda}=0.1$.]{
			\label{SNR_AER}
			\begin{minipage}{7.6cm}
				\includegraphics[width=\textwidth]{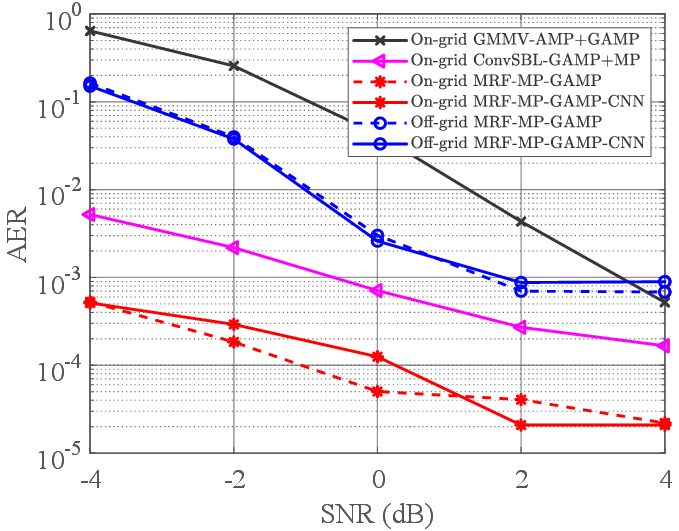} \\
				\vspace{-0.2cm}	
			\end{minipage}
		}
		\subfigure[Device identification performance under different deivce activities, where $\text{SNR}=2$ dB.]{
			\label{Activity_AER}
			\begin{minipage}{7.6cm}
				\includegraphics[width=\textwidth]{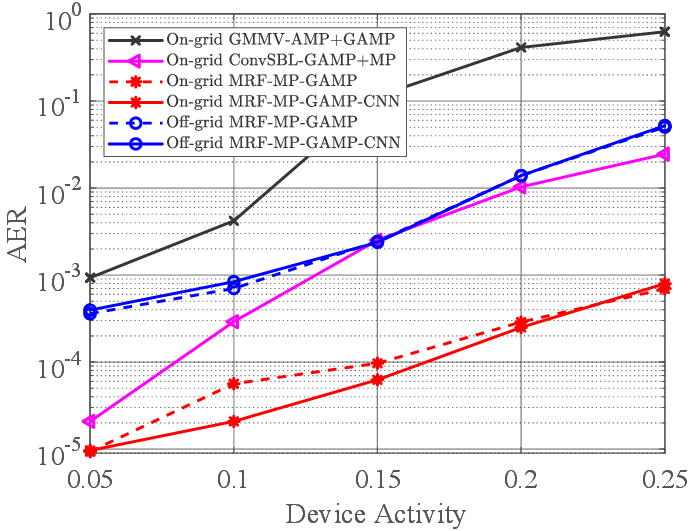} \\
				\vspace{-0.2cm}	
			\end{minipage}
		}
		\vspace{-0.2cm}
		\caption{Performance comparison for device identification among different schemes.} 
		\vspace{0.6cm}
		\label{DIPer}
	\end{figure}
	Finally, we show the performance comparison for device identification across various schemes in Fig. \ref{DIPer}, which plots the AER versus SNR and device activity, respectively. It is observed that the on-grid MRF-MP-GAMP and MRF-MP-GAMP-CNN deliver the best performance. Notably, when device activity exceeds 0.1, the off-grid MRF-MP-GAMP achieves performance comparable to that of the on-grid ConvSBL-GAMP+MP. This demonstrates the effectiveness of the proposed schemes. 
	
	\section{Conclusion}
	\label{Conclusion}
	This paper investigated grant-free random access tailored for LEO satellite communications in the absence of GNSS assistance. Exploiting the quasi-static property of the channel in the delay-Doppler domain, we proposed a spreading-based multi-frame OTFS transmission scheme to handle the large differential delay and Doppler shift. Then, the MP-based algorithm was designed for joint device identification, channel estimation, and signal detection, which is capable of eliminating the inter-user and inter-carrier interference and exploiting the 3D sparsity of the channel in the delay-Doppler-angle domain. Additionally, based on the statistical information provided by the receiver, the CNN detector was incorporated to improve the symbol detection performance for active devices. Simulation results demonstrated that the proposed multi-frame transmission scheme enhances system performance, and the designed algorithms outperform the conventional methods significantly in terms of the device identification, channel estimation, and symbol detection.  
	
	\appendices
	\section{Proof of Proposition 1}
	\label{Appendix A}
	We firstly focus on the input-output relationship for the single-input single-out (SISO), and hence omit the index of devices and antennas. The relationship for MIMO can be extended directly. According to (\ref{TSL})-(\ref{YTF}), we can derive the I/O in the time-frequency domain as 
	\begin{align}
		\label{a1}
		Y_q^{\text{TF}}&[n, m] =A_{g_{rx},r}(t, f) \mid_{t=(n+qN) T_{\text {sym}}, f=m \Delta f}  \nonumber \\ 
		& =\sum_{q^{\prime}=0}^{Q-1} \sum_{n^{\prime}=0}^{N-1} \sum_{m^{\prime}=0}^{M-1} X_{q^{\prime}}^{\text{TF}}[n^{\prime}, m^{\prime}] H_{q, n, m, q^{\prime}}[n^{\prime}, m^{\prime}],
	\end{align}
	where $H_{q, n, m, q^{\prime}}[n^{\prime}, m^{\prime}]=$
	\begin{align}
		&\sum_{i=1}^P h_i e^{\bar{\jmath} 2 \pi m^{\prime} \Delta f ([(n-n^{\prime})+(q-q^{\prime})_N] T_{\text {sym}}-\tau_i)}  e^{\bar{\jmath} 2 \pi(m-m^{\prime}) \Delta f T_{\text {sym}}} \times\nonumber \\
		& A_{g_{rx},g_{tx}}([(n-n^{\prime})+(q-q^{\prime})_N]T_{\text {sym}}-\tau_i,(m-m^{\prime})\Delta f-\nu_i) \nonumber \\
		&\times e^{\bar{\jmath} 2 \pi \nu_i((n+q N)T_{\text {sym}}-\tau_i)}.
	\end{align}
	Since the $T_{\text{cp}}>\tau_i$ and the adoption of rectangular waveform, $ A_{g_{rx},g_{tx}}([(n-n^{\prime})+(q-q^{\prime})_N]T_{\text {sym}}-\tau_i,(m-m^{\prime})\Delta f-\nu_i)\neq 0$ only if $n=n^{\prime}$ and $q=q^{\prime}$, i.e., there is no inter-symbol and inter-frame interference. Then, (\ref{a1}) can be simplified as
	\begin{align}
		\label{a2}
		Y_q^{\text{TF}}&[n, m]=\sum_{m^{\prime}=0}^{M-1} X_{q}^{\text{TF}}[n, m^{\prime}] H_{q, n, m}[n, m^{\prime}],
	\end{align}
	where
	\begin{align}
		\label{a3}
		H_{q, n, m}&[n, m^{\prime}]= 
		\frac{1}{T}\sum_{i=1}^{P} h_i \int_{T_{\text{cp}}-\tau_i}^{T_{\text {sym}}-\tau_i}
		e^{\bar{\jmath} 2 \pi \nu_i(t+(n+qN)T_{\text{sym}})}\nonumber \\
		&\times e^{-\bar{\jmath} 2 \pi m \Delta f(t+\tau_i-T_{\text{cp}})}
		e^{\bar{\jmath} 2 \pi m^{\prime} \Delta f(t-T_{\text{cp}})} dt.
	\end{align}
	We denote the length of CP as $M_{\text{cp}}=T_{{\text{cp}}}M\Delta f$. By substituting (\ref{xx}) and (\ref{a3}) into (\ref{a2}) and then applying the SFFT for $Y_q^{\text{TF}}[n, m]$, we can get the I/O in the delay-Doppler domain as
	\begin{align}
		&Y_q^{\text{DD}}[k, l]=\frac{1}{\sqrt{N M}} \sum_{n=0}^{N-1} \sum_{m=0}^{M-1} Y_q^{\text{TF}}[n, m] e^{\bar{\jmath} 2 \pi\left(\frac{m l}{M}-\frac{n k}{N}\right)}\nonumber \\
		&\overset{(a)}{=} \frac{1}{NM^2}\sum_{i=1}^{P} h_{i} \sum_{k^{\prime}}\sum_{l^{\prime}}X^{\text{DD}}_q[k^{\prime},l^{\prime}] \sum_{p=M_{\text{cp}}-(l_{i}+b_iM)}^{M+M_{\text{cp}}-1-(l_{i}+b_iM)} \nonumber \\ 
		&e^{\bar{\jmath} 2 \pi \tilde{k}_i q}e^{\bar{\jmath} 2 \pi \frac{p\left(k_i+\tilde{k}_i + d_iN\right)}{(M+M_{\text{cp}})N }} 
		\sum_{m=0}^{M-1} e^{-\bar{\jmath}2\pi\frac{ m(p-l+(l_i+b_iM)-M_{\text{cp}})}{M}} 
		\nonumber \\
		&\sum_{m^{\prime}=0}^{M-1} e^{\bar{\jmath}2\pi\frac{ m^{\prime}(p-l^{\prime}-M_{\text{cp}})}{M}}\sum_{n=0}^{N-1} e^{-\bar{\jmath}2\pi\frac{ n(k-k^{\prime}-k_i-\tilde{k}_i-d_iN)}{N}} \nonumber \\
		&=\frac{1}{\sqrt N}\sum_{i=1}^{P} h_{i} \sum_{k^{\prime}}\sum_{l^{\prime}}X^{\text{DD}}_q[k^{\prime},l^{\prime}] \Pi_N(k-k^{\prime}-k_i-\tilde{k}_i) \nonumber \\  &\sum_{p=M_{\text{cp}}-(l_{i}+b_iM)}^{M+M_{\text{cp}}-1-(l_{i}+b_iM)} e^{\bar{\jmath} 2 \pi \frac{p\left(k_i+\tilde{k}_i + d_iN\right)}{(M+M_{\text{cp}})N }} \delta((p-l^{\prime}-M_{\text{cp}})_M) \nonumber \\
		&\times\delta((p-l+l_i+b_iM-M_{\text{cp}})_M) e^{\bar{\jmath} 2 \pi\tilde{k}_i q}  \nonumber \\
		&=\frac{1}{\sqrt N}\sum_{i=1}^{P} h_{i} \sum_{k^{\prime}}\sum_{l^{\prime}}X^{\text{DD}}_q[\left\langle k-k^{\prime}\right\rangle_{N},\left(l-l^{\prime}\right)_{M}]
		e^{\bar{\jmath} 2 \pi\tilde{k}_i q}  \nonumber \\
		&\quad \times \Pi_N(k^{\prime}-k_i-\tilde{k}_i) e^{\bar{\jmath} 2 \pi \frac{\left(k_i+\tilde{k}_i + d_iN\right)}{(M+M_{\text{cp}})N }(M_{\text{cp}}+l-l_{i}-b_iM)}
		\nonumber \\
		&=\sum_{l^{\prime}=0}^{M-1} \sum_{k^{\prime}=\lceil-N / 2\rceil}^{\lceil N / 2\rceil-1} H^{\mathrm{DD}}\left[k^{\prime}, l^{\prime}, l\right] \phi_{q}[l^{\prime}] \nonumber\\
		&\qquad \qquad\qquad\qquad\times  X_q^{\text{DD}}\left[\left\langle k-k^{\prime}\right\rangle_{N},\left(l-l^{\prime}\right)_{M}\right],
	\end{align}
	where $(a)$ is obtained by utilizing (\ref{tap}) and $\phi_{q}[l^{\prime}]$ has the similar form as (\ref{phiq}). Let $\bar h_{i}=h_{i}e^{\bar{\jmath} 2 \pi\nu_i(M_{\text{cp}}T_s-\tau_i)}$, and then we have
	\begin{align}
		\label{a4}
		H^{\mathrm{DD}}\left[k^{\prime}, l^{\prime}, l\right]=\frac{1}{\sqrt N} \sum_{i=1}^{P}& \bar h_{i} e^{\bar{\jmath} 2 \pi \nu_{i}l T_s} \delta\left(l^{\prime} T_{\mathrm{s}}-(\tau_{i})_T\right) \nonumber \\ 
		&\times \Pi_N(k^{\prime}-NT_{\text{sym}}\nu_{i}). 
	\end{align}
	Multiplying (\ref{a4}) by the antenna phase rotation $e^{\bar{\jmath} \pi n_z \Theta_{u}^z}$ and $e^{\bar{\jmath} \pi n_y \Theta_{u}^y}$, we get the result in Proposition 1.
	\bibliographystyle{IEEEtran}
	\bibliography{bibfile}
\end{document}